\documentclass[preprint,12pt]{elsarticle}

\usepackage{amssymb}
\usepackage{enumitem}
\usepackage[dvipsnames]{xcolor}
\usepackage{float}

\begin{document}

\begin{frontmatter}

\title{A numerical toy model of Langevin dynamics provides real-time visualization of colloidal microdroplet evaporation}

   \author[IFPAN]{G. Derkachov \corref{cor1}}
   \author[ICCEUW]{T. Jakubczyk}
   \author[IFPAN,FU]{S. Alikhanzadeh-Arani}
   \author[IFPAN,MAGTOP]{T. Wojciechowski}
   \author[IFPAN]{D. Jakubczyk}
   \cortext[cor1]{G. Derkachov, derkaczg@ifpan.edu.pl}
 \address[IFPAN]{Institute of Physics, Polish Academy of Sciences,\\
             al.~Lotnik\'{o}w 32/46, Warsaw, PL-02668, Poland}
\address[ICCEUW]{Institute of Control and Computation Engineering, Warsaw University of Technology,\\
             ul.~Nowowiejska 15/19, Warsaw, PL-00665, Poland}
\address[FU]{Present address : Frahangian University, P.O. Box 14665-889, Tehran, Iran}
\address[MAGTOP]{International Research Centre MagTop,\\
             al.~Lotnik\'{o}w 32/46, Warsaw, PL-02668, Poland}

\begin{abstract}
We have developed and tested a simplified but versatile numerical model of nanoparticles' aggregation using Langevin dynamics. The model is particularly capable of simulating aggregation in an evaporating (or condensing) microdroplet. It runs on a graphics processing unit (GPU), which makes it sufficiently fast for real-time conceptualization tasks. We have verified the results of modeling against the findings from two types of experiments we conducted in electrodynamic traps. Firstly, our model helped us to elucidate the phenomenon of scattering `revival', often observed during the evaporation of composite microdroplets. Further on, we were able to mimic our experiments, in which the microdroplets were dried up to form nanoparticle (NP) aggregates, and then soft-landed. Thus we could compare model predictions with SEM imaging. The model was tested for up to $2.5\times 10^5$ nanoparticles of several coexisting types. Several types of interactions can be accounted for: inter-particle: Lennard-Jones and Coulomb; external: dispersion medium viscosity, centrifugal force, gravity, surface tension, and interface movement. Brownian motion of nanoparticles can be freely controlled. The core program is accompanied by scripts extracting statistical NP aggregates properties in post-processing -- fractal dimension and radial distribution functions. The codes are made available in public repositories. Several diverse evolution scenarios are presented.

\end{abstract} 

\journal{\empty}
\end{frontmatter}


\section{Introduction}
It has been well-known that atmospheric models are multi-scale \cite{tao2009multiscale,byun1999science} In particular, of course, models of clouds are such and require input at the micro-scale \cite{pruppacher1998microphysics, majeed2001microphysics}. For instance, the energy transfer in a single droplet is fundamental to the energy flows in the cloud, and the optical properties of a single microdroplet strongly influence the optical properties of the cloud (like albedo). Also, the content of solid nano- and microparticles in atmospheric droplets can be significant and can have a fundamental impact on the microscale properties and processes, which are inherited at higher scales \cite{pani2023aerosol}. 

Atmospheric droplets can acquire solid particles via diverse processes, like nucleation scavenging and/or impaction scavenging (see e.g. \cite{pruppacher1998microphysics}). In a clean environment, the solid particles are rather sparse in the atmospheric droplets. Still, fog droplets in dusty or polluted environments can be rich in solid nano- and microparticles \cite{ogren1992measurements, gonzales1977electron,li2013microscopic}. 

In order to properly parametrize the microscopic properties of the larger system, a good understanding of its micro-scale processes and properties is required. Thus, there is a demand for micro-scale modelling on the one hand, and for fast and accurate remote diagnostic methods operating at micro-scale. Optical methods seem to suit very well and indeed they span various applications, from atmospheric physics and chemistry \cite{singh2022quantifying, kocifaj2022diffuse, brown2012review, xie2014study} through ecological issues \cite{jiaqiang2022soot},   to industrial processes \cite{santosIndustrialApplicationsNanoparticles2015, KUMAR2021109042, JAFARIDAGHLIANSOFLA2018559}. There is also a wide range of microphysical modelling methods (which we discuss later on).

Our group focuses on developing methods for the optical characterization of evaporating microdroplets containing diverse nanoparticles (NPs). The task, if approached straightforwardly, would involve solving the inverse (scattering) problem, which is ill-conditioned, ambiguous, and generally hopelessly complex. It is well-established that the presence of dispersed phase \cite{sorensen2001light, xu2015light}, especially the structures (aggregates) it forms within the microdroplet volume, significantly influences both its optical properties and the evaporation process. It would then be constructive to know something about possible internal structures and scenarios of their evolution. Even though theories and models of evaporation and light scattering have been developed for over a century, our understanding of such complex systems is limited. As far as the evaporation thermodynamics of a microdroplet of suspension is concerned, several questions persist. For instance, what are the limits of applicability of the kinetic and the continuous flow theories for the given stage of a complex microdroplet evolution (compare \cite{holyst2013evaporation} for pure liquid microdroplets)? As the internal structures grow, they reduce the volume accessible to liquid molecules and the open surface area available for evaporation. What is the internal structures' influence on the diffusion process of liquid molecules to the microdroplet's surface?
	In (our) experimental investigations of evaporating microdroplets of suspension, an irreplaceable role is played by traps -- electrodynamic or acoustic, which keep the microdroplets levitating, preventing the influence of substrates or suspension threads. We infer the parameters of an evaporating microdroplet from the properties of scattered coherent light and by electrostatic weighting. Estimating the microdroplet size from static light scattering involves comparing the angle-dependent intensity of scattered light with Mie theory predictions \cite{jakubczykCombiningWeightingScatterometry2013}. However, this is possible only at low concentrations of NPs, since the theory assumes a homogeneous sphere. The dynamics of internal and surface structure formation could be inferred using dynamic light scattering or speckle movement analysis. However, capturing the image of the internal/surface structure in a levitating microdroplet is impossible so far. SEM images are only obtainable at the final stage of evolution after drying up the microdroplet and soft-landing it on the substrate \cite{wozniak2015formation, archer2019sodium}. Again, it becomes obvious that a comprehensive understanding of the entire process necessitates some insights into the dynamics of the nanoparticle aggregation process, which an aggregation model can provide.

\subsection{Motivation}
\begin{figure}[htb]
\centering
    \includegraphics[scale=0.5]{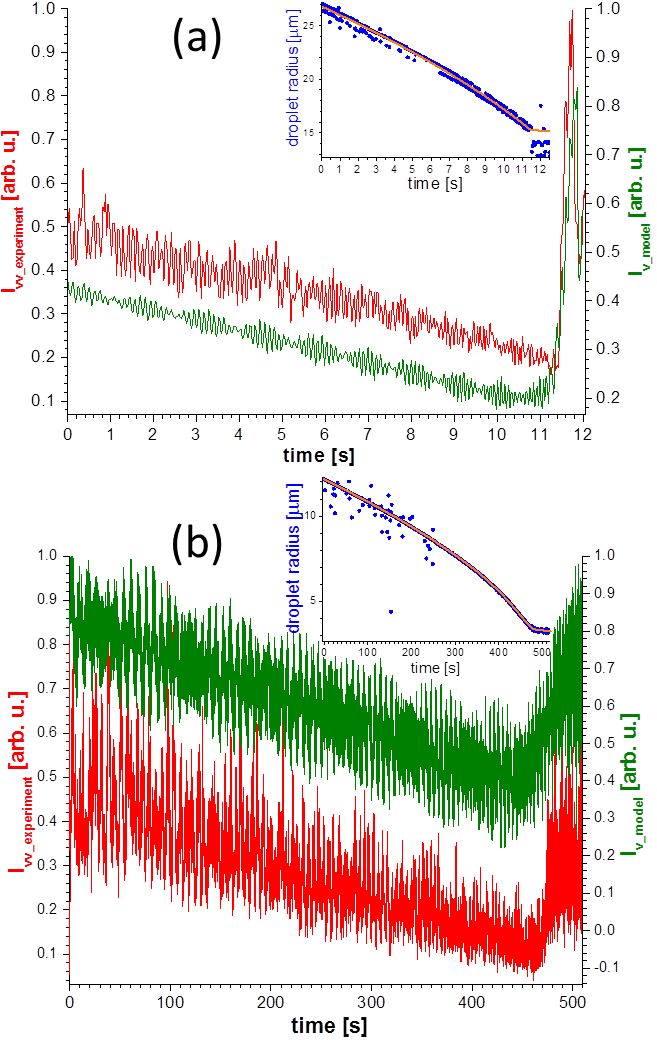}
    \caption{Two cases of scattering `revival' from single evaporating microdroplets of suspension in an electrodynamic trap: panel (a) -- 450-nm-diameter SiO$_2$ NPs in dimethyl sulfoxide (DMSO); panel (b) -- 200-nm-diameter polystyrene (PS) NPs in diethylene glycol (DEG). Red traces -- experimental light intensity integrated over the field-of-view; green traces -- corresponding modeling results obtained with Eqn. \ref{tot_scat}. Radii evolutions are shown in corresponding insets: blue dots -- experimental results obtained with optical method and electrostatic weighting (see \cite{jakubczykCombiningWeightingScatterometry2013}); orange line -- radius evolution used for Mie theory calculations and droplet evaporation modeling.}
    \label{fig:Intensity_revival}
\end{figure}

 A specific motivation is associated with the challenging scenarios of scattered light intensity evolution that we often encounter in our experiments with microdroplets of suspension. Two examples are shown in Fig. \ref{fig:Intensity_revival}. After an initial decrease in the intensity of the scattered light, which roughly follows a reduction in the geometric cross-section of the microdroplet, a very significant `revival' occurs, associated with the appearance of the suspended phase on the surface of the microdroplet.  Proper handling of such cases is important for the accurate prediction of albedo \cite{dey2024using,pani2023aerosol,chen2023influence}  and refractive properties of the atmosphere \cite{herman1975effect,takemura2002single,chen2024surface}, as well as application of the lidar equation (see e.g.:  \cite{weitkamp2005range,kong2022lidar,li2020retrieval}). It is closely related to the well-known radar echo artefact caused by the chaff cloud (see e.g. \cite{garbacz1978chaff}).
 The solid particle content in the microdroplet can be significant, as high 
 Using just the droplet cross-section, calculated e.g. with Mie theory for constant (effective) parameters, to estimate the scattering of light, would severely underestimate its intensity.  As the first approximation, it can be augmented with the formalism proposed e.g. in \cite{sorensen2001light} (keeping in mind that it was devised for fractal aggregates in free space rather than droplets acting as spherical resonators). In the far field, the total scattered light intensity $I_{\rm scat}$ would come then from the light scattered by an equivalent homogeneous sphere with effective parameters $I_{\rm Mie}$ and the light scattered coherently from individual NPs:
 \begin{equation}
     I_{\rm scat}\simeq I_{\rm Mie}+\alpha N_{\rm surf}^2 \mbox{ ,}
     \label{tot_scat}
 \end{equation}
where $N_{\rm surf}$ is the total number of NPs at the droplet surface (where the light field is most intense), and the factor $\alpha$ may depend, among others, on the droplet and NP radii and the NP structure's mass and surface fractal dimensions. $N_{\rm surf}$ and (eventually) the corresponding fractal dimensions must be provided by a model. Since the scattering revival phenomenon seems quite general, we expect that it is caused by substantial changes in the composite droplet structure, rather than by a minute structure alteration. Accordingly, a simple and robust model should be sufficient initially. Even a simple numerical model serves as a virtual laboratory, expediting hypothesis testing and providing rapid feedback on method viability.
In this work, we investigate the applicability of a fairly simple numerical 'toy' model of nanoparticle aggregation in an evaporating microdroplet, that we developed. A ballpark elucidation of the intensity revival phenomenon, using Eqn. \ref{tot_scat} together with the model, is presented in Fig. \ref{fig:Intensity_revival}. We were also able to visualize different scenarios of microdroplet evolution beyond the specific motivation presented in this section.

\subsection{The requirements of the toy model}
 The model needs to predict the trajectories of NPs over time, describing the dynamic process within evaporating/condensing microdroplets. In our experiments, the microdroplet size varies from about a hundred micrometers down to hundreds of nanometers, with the dispersed phase NPs being spheres ranging from $\sim10$ nanometers to $\sim1$ micrometer in diameter. These size ranges match those encountered in cloud and fog droplets quite well. However, NPs encountered in the atmosphere are obviously generally not spherical. Some atmospheric NPs, such as those composed of silica, acquire an electrical charge when mixed with water, causing them to repel each other. But, of course, their proximity causes strong attraction and bonding.
	Moreover, NPs interact with the microdroplet surface, unable to penetrate it due to surface tension forces. The microdroplet experiences Brownian motion as a whole, and each NP is influenced by Brownian forces. However, the movement of NPs is relatively slow (the whole process of drying up the microdroplet takes several minutes to hours) and hydrodynamic forces can be considered negligible. Under specific conditions, such as a strong electric field in the trap (creating a deep potential well), the Brownian motion of the entire microdroplet may be constrained to Brownian rotation, when trapping potential restricts microdroplet lateral movement but allows rotation. As a rough approximation, we characterize the average centrifugal force acting radially (in all directions) directly on each NP due to Brownian rotation, with a single scalar, which is (somewhat artificially) linked to an average scalar rotation speed. Thus, the movement of the NP is not transmitted from the Brownian rotation of the entire microdroplet through drag forces. Considering the density difference between the liquid and the nanoparticle material, gravitational/Archimedean forces must also be considered.	

\subsection{Selecting the nanoparticle aggregation scenario}
In the literature, models or numerical procedures capable of generating the final aggregate shape present at least two different approaches: one uses a purely "geometrical" approach to NPs and their trajectory and can be coded into a straightforward numerical algorithm capable of generating the aggregate without delving into the specifics of physical forces; the other relays on physical interactions and the NP's position/trajectory is calculated from the proper force fields.  The model of the first type can predict the final form of the aggregate but cannot predict the dynamics of its formation. The other is capable of both but is significantly more resource-demanding.	In the first group of algorithms/models, we can include diffusion-limited aggregation (DLA)\cite{witten1981diffusion,l.m.sanderDiffusionlimitedAggregationKinetic2000}, DLCA \cite{kolbScalingKineticallyGrowing1983,meakinBallisticDepositionSurfaces1986}, reaction-limited aggregation (RLA), and RLCA \cite{meakinStructureKineticsReactionlimited1988}.  The second group is exemplified by molecular dynamics (MD) \cite{ulbergMolecularDynamicsSimulation1993}, Brownian dynamics (BD)\cite{markutsyaBrownianDynamicsSimulation2008} and Langevin/Stokesian dynamics (LD) \cite{garcia-palaciosLangevindynamicsStudyDynamical1998}, or computational fluid dynamics (CFD; e.g. lattice Boltzmann methods (LBM))\cite{wendtComputationalFluidDynamics2008}, though the nomenclature is somewhat diffuse in this field. There is a multitude of available codes/packages implementing the above methods, in particular of the second group. Some packages ((like LAMMPS or OpenMM)) are more flexible regarding force field selection. Others help to set up the numerical experiment (Pizza.py, ParmEd) and have features like GUI, results visualization, etc.
	For our purposes, we need the model of the second kind, operating at nano- and microscale rather than at the molecular level. We've decided that a model based on Langevin dynamics would be most convenient for our task - the Langevin equation combines deterministic forces (potential energy gradients) with random forces representing thermal fluctuations. The simulation involves updating the position and velocity of the NP at each time step, incorporating the deterministic and stochastic components. Numerical integration methods, such as the Euler method, Verlet algorithm (and its derivatives like leapfrog integration), or more sophisticated schemes like the stochastic Runge-Kutta methods, can be employed to simulate the Langevin dynamics. The choice of integration method can impact the accuracy and stability of the simulation. It's worth noting that the Langevin dynamics simulation is a computationally intensive process, and careful consideration should be given to the choice of the numerical method, time step, and simulation parameters to ensure accurate and meaningful results. When selecting a software vehicle for our toy model, we were aware that some of the advantages of the full-fledged LD would have to be sacrificed.

\section{The toy model} \label{The_model}
\subsection{The main features and concepts of the algorithm \label{concepts_of_the_algorithm}}
	In the presented work, we adapted the solver from Nvidia \cite{green2010particle}, which is optimized for parallel calculations on Nvidia GPUs. Though designed with much-simplified physics, it turned out to be fairly flexible in terms of the force field used, while it has an inherently built-in visualization feature, also with a simple GUI, which was very attractive for our applications. The current model builds on the experience gained from using our previous model written in Matlab/Simulink (compare e.g. 	\cite{derkachov2008drying, wozniak2015formation}; the code was published on GitHub \cite{MatlabSimulinkAggregation}). It must be kept in mind, that the presented model does not pretend to cover all aspects of physical reality -- it is a conceptualisation tool to roughly test some hypotheses.
	
	The previous code helped us to understand the surface thermodynamics in evaporating microdroplets of suspension but had serious limitations (limited number and type of NPs, type of interactions, speed). In the current model, these limitations were significantly relaxed -- the C++/nvc code running on GPU (Tesla K40) is several orders of magnitude faster while allowing for a few hundred thousand nanoparticles of several types. While following the evolution of a 20-micrometer droplet with 5000 NPs required days, in the case of the older code, it requires only several seconds with the newer one. It is thus perceived by the user as real-time. Several types of interactions are included and further can be easily added. As it turns out, such a model is also capable of handling many scenarios in bulk suspension, and possibly also the deposition of microdroplets/aggregates on the surface. It has been complemented with a set of Matlab scripts that allow extraction, in post-processing mode, such information as the evolution of the fractal dimension, the number of sub-aggregates, surface pressure, aggregate statistics, etc. The codes are made available in the public repository \cite{githubGPUparticles}. The current version of the code has provided us with the first insights into the `doughnut' structure formation, which we sometimes (rarely) observed in our linear trap (see \ref{fig:Oponka_SEM&Simulation} below).
In comparison to the original codes from Nvidia \cite{CUDA,green2010particle}, which are designed primarily for computer games, we've tried to implement more (but not entirely) rigorous physics. The model considers the NP system to be close to equilibrium. A given number of NPs with desired parameters (radius, mass, charge) are initially randomly distributed in a sphere -- a microdroplet of a given (initial) radius. Several types of NPs can be used simultaneously. The radial non-equilibrated forces are introduced to the system by imposing the irreversibly moving surface of the evaporating droplet (gas-liquid interface) dragging up the NPs dispersed in evaporating liquid volume. The interaction of an NP with the surface (surface tension) was described as a spring with the equilibrium point at the distance of an NP radius from the surface (the equilibrium point can be easily changed in the code). The spring constant can be set separately for each NP type. The microdroplet evaporation/condensation proceeds following the radius-square-law \cite{jakubczyk2012evaporation} and the evaporation/condensation rate can be freely defined. It must be underlined here that the interaction of NPs with the surface and/or surface movement can be easily switched off to mimic the conditions of NPs self-assembly in the bulk liquid. Two inter-particle interaction types were accounted for (so far): the effective Lenard-Jones (LJ) potential and long-range electrostatic interactions. The interactions within each NP type and between types can be freely set. Both LJ and electrostatic interaction ranges were limited, but their values were not shifted. The gravitation and centrifugal force associated with the microdroplet rotation were also introduced. Further forces can be easily added to the code. There is also a selectable option of imposing Brownian NPs motion. It is obtained by adding a randomly directed vector, whose magnitude can be controlled, to the calculated velocity. Movements of each NP inside the evaporating microdroplet were simulated with the classical Newton's equation of motion. The (numerical) stability of NPs collisions, which at finite time-step can yield non-physically high velocities, was taken care of by introducing a dissipative force in the form of non-physically high liquid viscosity, which however conforms to Stokes' law.

\subsection{The outline of the algorithm.}
The core algorithm's main steps (loop) are as follows:
\begin{enumerate}[label=(\roman*)]
\item Arrange NPs into the cells (by second indexing), according to their position, to identify which are close to each other and can interact. In the described simulations, the interaction range was set to $4\left(r_{\rm NP(A)} + r_{\rm NP(B)}\right)$, where $r_{\rm NP(..)}$ are the radii of the colliding NPs (A and B). This range can be changed in the code.
\item For NPs that are in the range of interaction potential (Lennard-Jones, Coulomb) calculate the forces. LJ and ‘pure' Coulomb potential was truncated without shifting. 
\item Calculate the new microdroplet radius.
\item For NPs that are in the range $(-r_{\rm NP}, r_{\rm NP})$ from the droplet surface, calculate the forces and store them in the table.
\item Add forces of gravity and (effective) viscosity. Other forces of NP interactions with external/non-local fields (such as centrifugal force) are added here.
\item Calculate the new velocity and position for all NPs. Brownian motion is emulated here by adding a random velocity vector (with a controlled magnitude). The notion of temperature is not included in the model, but the magnitude of the random velocity component (passed as a parameter) imposes it externally.
\end{enumerate}
The simulation results are displayed online and can also be recorded in a binary file for viewing and post-processing. The simulation can also be stopped and restarted with different parameters from the saved NPs configuration. Online previewing is used for preliminary hypotheses and parameters testing with a larger time-step, while in the proper numerical experiment, a fine time-step is used to reduce harmful numerical effects, while only every $j$th "frame" is recorded.
\subsection{Verification of the model versus SEM images} \label{model_vs_SEM}
As mentioned above, we carried out a series of experiments in a linear quadrupole trap, in which microdroplets with NPs were dried to form dry NPs aggregates, and soft-landed (deposited on a substrate) at different stages of evolution (drying) \cite{wozniak2015formation,archer2019sodium}. In this work, we used only the SEM images of fully dried microdroplets, which retained their shape after soft-landing. Thus, we could unambiguously compare at least the final stage of evolution obtained from both the experiment and the model.

\begin{table}[t]
\caption{Main parameters of selected simulations. $R_0$ is the initial microdroplet radius, $A$ is the evaporation rate, $\epsilon$ is the dispersion energy,  $\omega$ is the effective rotation rate, $r_{\rm NP}$ is the NP's radius.} \label{sim_parameters}
\begin{center}
\begin{tabular}{cccccccc}
\hline\hline
Fig. \# & $R_0$ & A & $\epsilon$ &   $\omega$ & $r_{\rm NP}$ & \# of NPs & NP charge\\
    & $\left[\mu {\rm m}\right]$ & $\left[\mu {\rm m/s}\right]$ & [J] &   [rad/s]& $\left[\mu {\rm m}\right]$ & & [arb. u.]\\
\hline
1 & 20 & 0 & 0.023 & 1 & 0.5 & 10 000 & 0 \\
\hline
3 & 20 & 0 & 0.023 & 0 & 0.5 & $\begin{array}{cc}
									10 000\\ 
									10 000 
								\end{array}$ & $\begin{array}{cc}
													1 \\
													-1
												\end{array}$\\
3 & 20 & 0   & 0.023 & 1 & 0.125 & 10 000 & 0 \\
5 & 20 & 3.5 & 0.023 & 0 & 0.125 & 10 000 & 0 \\
8 & 60 & 5   & $0.23 \cdot \left[ \begin{array}{cc}
									0.155 & 0.155 \\
									0 & 1.26
								  \end{array} \right]$ & 1 & $\begin{array}{cc}
								  								0.25 \\
								  								1.5
								  							  \end{array}$ & $\begin{array}{cc}
	39500\\
	1500							  							  
\end{array}$ & 0 \\						  							  					
\end{tabular}
\end{center}
\end{table}
\section{Microdroplet evolution analysis}
\begin{figure}[htb]
\centering
    \includegraphics[scale=0.05]{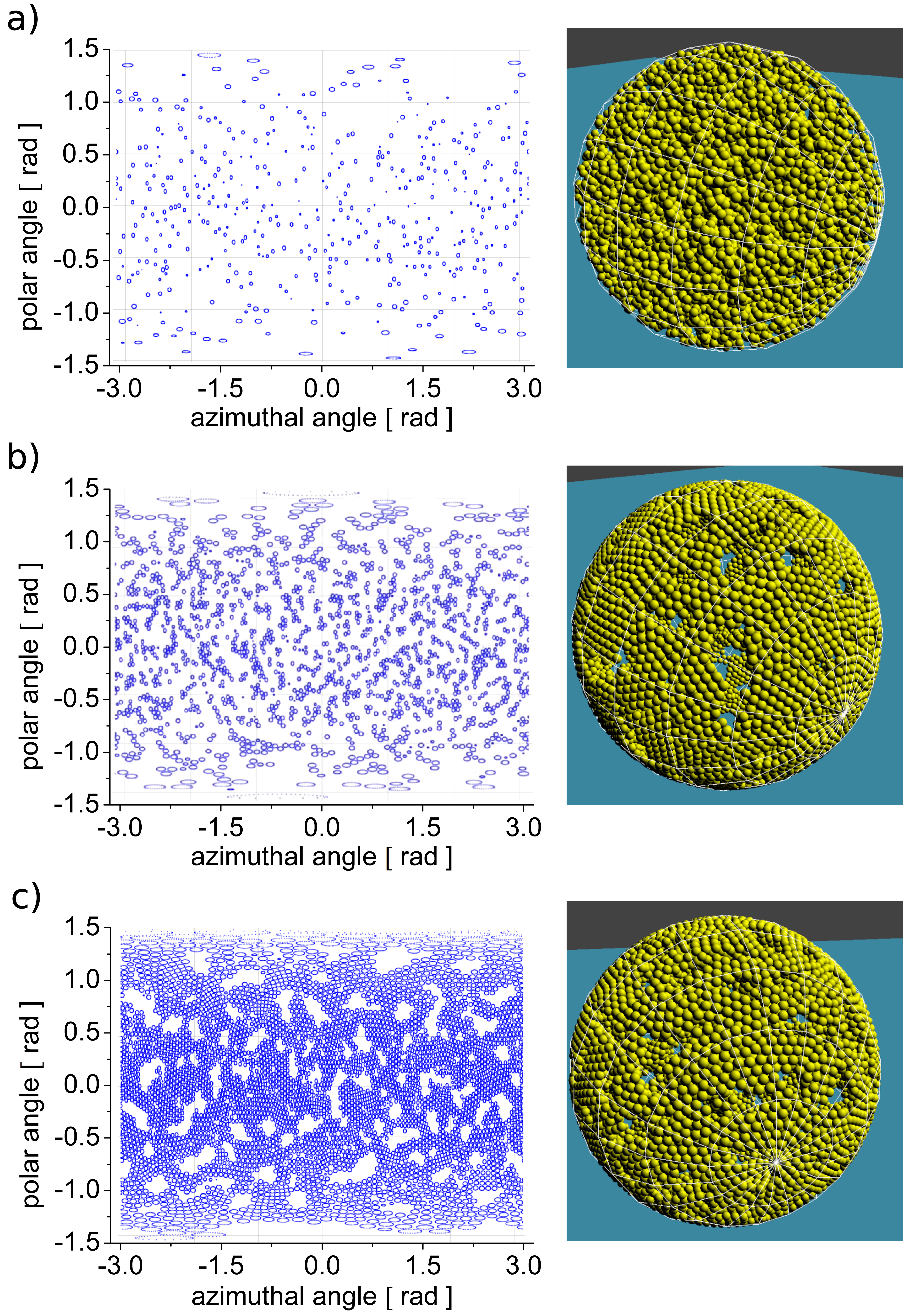}
    \caption{Visualization of the formation of a shell of NPs in a microdroplet of suspension. Figure rows: a) early, b) intermediate, and c) late stage of the evolution respectively. Right panel: a perspective view of the evolving microdroplet. Left panel: projection of the microdroplet surface onto a plane. Blue circles/ellipses (circles elongated due to the projection) correspond to the NPs intersections with the microdroplet surface.  Specific model parameters: The microdroplet wasn't evaporating -- had the constant radius $R_0=20$ $\mu$m, but underwent random Brownian rotations (tumbling) generating an (effective) isotropic centrifugal force, which would formally correspond to the angular frequency $\omega$ = 1 rad/s. There were $10^4$ NPs with a radius of 500 nm. The NPs were electrically neutral and the NP-NP interaction was described by the Lennard-Jones potential.}
    \label{fig:Aggregate_evolution}
\end{figure}

We've conducted a series of numerical experiments testing the model's capabilities. The parameters of several of these discussed here, are given in Table \ref{sim_parameters}. It must be underlined that, as far as possible, the parameters corresponded to those encountered for the atmospheric droplets (aerosols), like microdroplet and NP radii and material densities, NPs number, and microdroplet rotation and evaporation rates, or were derived from the corresponding material properties, like LJ parameters. But simultaneously it must be admitted that NP charge and surface tension were rather arbitrarily chosen, while the viscosity of the host liquid was non-physical since it served as a numerically stabilising factor, as discussed in subsection \ref{concepts_of_the_algorithm}. The visualizations of results are presented in corresponding figures. As a result of the simulation, we obtain the temporal evolution of the NPs' positions. For example, in Fig. \ref{fig:Aggregate_evolution}, we show raw visual results of the simulation, accompanied by the results of elementary post-processing, of the NP structure (aggregate) formation in a non-evaporating tumbling microdroplet of suspension. Three distinctive stages of evolution are shown: (a) early, (b) intermediate, and (c) late. In the right panel, a perspective view of the evolving microdroplet is shown, while in the left panel, the projection of the microdroplet surface onto a plane is visualized. The temporal evolution of the NPs' positions provides a complete description of the aggregation dynamics. In this form, however, it is hardly informative, so a more comprehensible description is needed. Suitable numerical procedures were then constructed (MATLAB scripts) to obtain the desired statistical parameters of the microdroplet/aggregate evolution.
       
With the task of remote optical characterization of NP aggregates in mind, we chose the two most widely used complementary parameters that describe the state of an aggregate on global and local scales, respectively -- fractal dimension \cite{ferri2022MassFractal} and radial distribution function (RDF) \cite{larsen2018methodComputingRDF_Cloud}. They were shown to correlate with the properties of the scattering images \cite{onofriOpticalCharacterisationNanoparticle2011,jungblutModelingNanoparticleAggregation2019}. In particular, the counterpart of RDF in Fourier space  -- the structure factor, can be measured in scattering experiments.
                
To extract the fractal dimension and the RDF from the position of NPs, we followed the procedure that was described in \cite{ferri2022MassFractal,kopera2018computingRDFof_Particles}.  
For convenience, here is a concise outline of the core concept.
\subsection{Fractal dimension}
Fractal parameters are standard mathematical constructs for describing self-similar objects. In the case of real aggregates, we encounter notions of mass-fractal and surface-fractal. In the case of a mass-fractal, we can estimate the mass (or the number of elementary building blocks -- primary particles -- NPs in our case) of the aggregate, knowing the size of its bounding box and the space dimensionality. While the relationship between the mass and size of a solid is $m \sim {d}^3$, where $d$ is the characteristic size of the solid, for a mass-fractal, it becomes $m \sim {d_{\rm NP}}^{D_{\rm f}}$, where $d_{\rm NP}$ is the mean diameter of primary particles (NPs) and $D_{\rm f}$ is the fractal dimension. If an aggregate consists of $N$ identical spherical NPs of the diameter $d_{\rm p}$, we can link the number of nanoparticles $N$ to the radius of gyration $R_{\rm g}$ and obtain the so-called "fractal equation":
\begin{equation}
    N = k_{\rm f}\left( \frac{2R_{\rm g}}{d_{\rm NP}} \right)^{D_{\rm f}} \mbox{,}
    \label{N}
\end{equation}
\begin{equation}
    R_{\rm g} = \sqrt{\frac{1}{N}\sum_{n = 1}^{N}\left( \vec{r_0} - \vec{r_n} \right)^2} \mbox{ ,}
    \label{Rg}
\end{equation}
\begin{equation}
    \vec{r_0} = \frac{1}{N}\sum_{n = 1}^{N}\vec{r_n} \mbox{ ,}
    \label{r0}
\end{equation}
where $R_{\rm g}$ is measured from the center of mass $\vec{r_0}$, $\vec{r_n}$ is the position of $n$th NP and $k_{\rm f}$ is the fractal prefactor (structural coefficient) \cite{lapuertaGeometricalDeterminationLacunarity2010}. The value of this constant is widely debated, and different considerations yield values from 1 to $\sim 6$. For an aggregate with a compact hexagonal structure, the value of $k_{\rm f}$ can be determined as 1.593 (see e.g. \cite{mroczka2012AlgorithmsMethodsOpticalStructureFactor}). Since, in our system/simulation, the formation of such an aggregate is allowed, we used the above value of $k_{\rm f}$. In each simulation, we follow the evolution of a single aggregate of $N$ NPs (in particular, from the moment when a single, droplet-filling aggregate is formed) -- the shrinking surface of the evaporating droplet compresses the fractal aggregate and $R_{\rm g}$ decreases for constant $N$. Since we know the number and position of participating NPs, we can solve the fractal equation for $D_{\rm f}$.

As mentioned earlier, the light scattering on a fractal is linked to the above-considered parameters. For instance, for parameters encountered in atmospheric research (say, $d_{\rm NP}=200$~nm, light wavelength $\lambda=500$~nm), the second term of equation \ref{tot_scat} can be expressed following \cite{sorensen2001light} as:
\begin{equation}
    I_{\rm NPscat}(q) \propto N_{\rm surf}^2\left( qR_{\rm g}\right)^{-D_{\rm f}} \mbox{,}
    \label{NPscat}
\end{equation}
where $q=4\pi\sin (\Theta/2)/\lambda$ is the scattering wave vector and $\Theta$ is the scattering angle.

\begin{figure}[htb]
    \centering
    \includegraphics[scale=0.5]{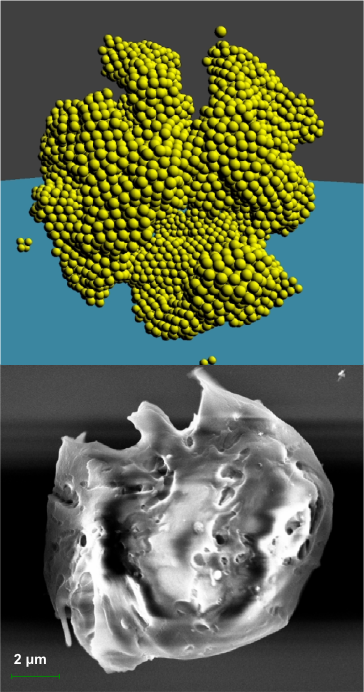}
    \caption{Examples of NP aggregates of mildly fractal structure. Lower panel -- SEM image of an NP aggregate we obtained from drying a free microdroplet of an aqueous suspension of 100-nm-radius SiO$_2$ NPs with an addition of sodium dodecyl sulfate (100 mM initially) in a linear electrodynamic quadrupole trap, for 25 min. Upper panel -- the result of the simulation for model parameters corresponding to the red curve in Fig. \ref{fig:Df_evolution} and radial distribution functions in Fig. \ref{fig:RDF_Lennard_Jones}.}
    \label{fig:LJ_only_SEM&Simulation}
\end{figure}

\begin{figure}[htb]
    \centering
    \includegraphics[scale=0.5]{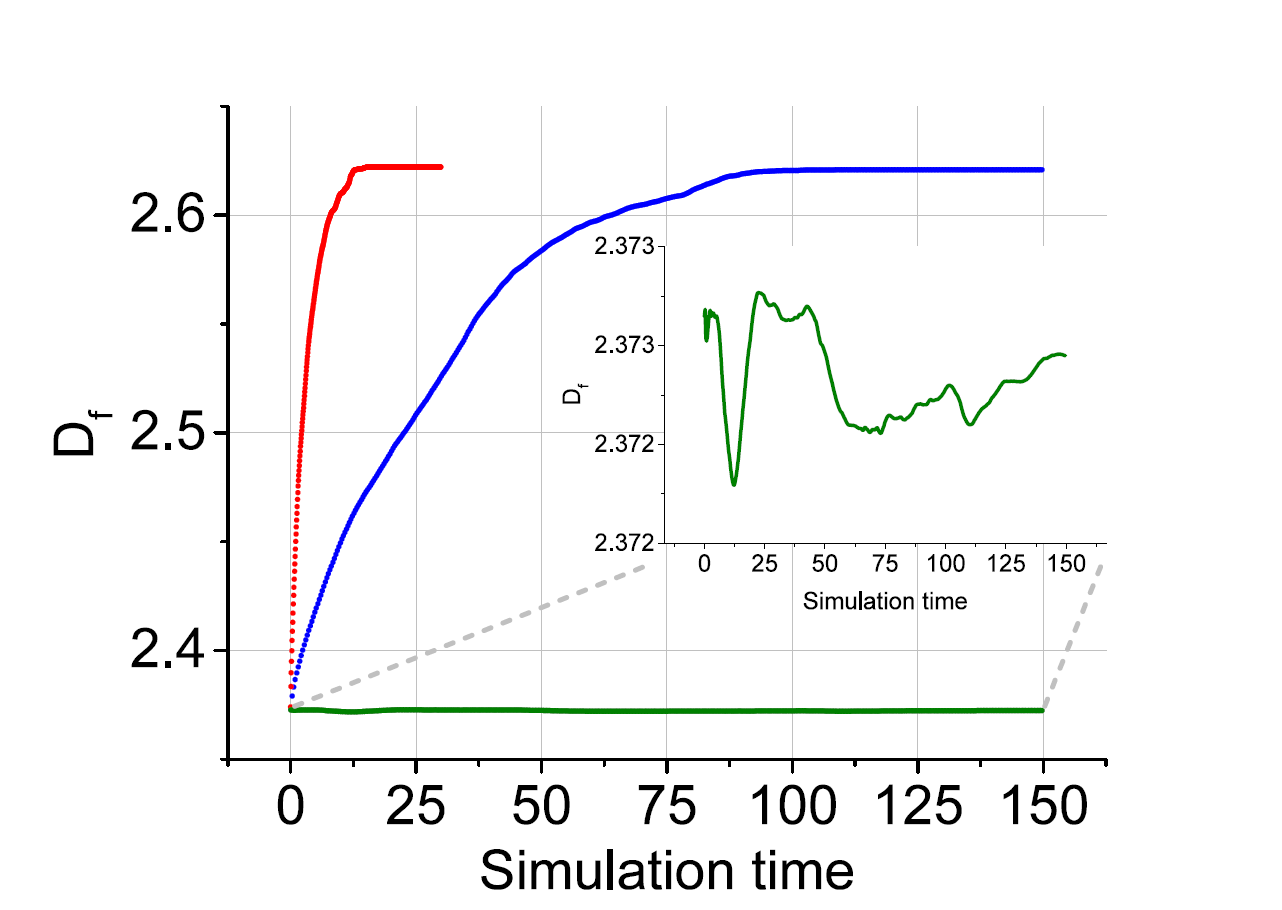} 
    \caption{Examples of temporal evolutions of the fractal dimension of aggregates forming in a bulk liquid (mimicked by a stationary microdroplet remaining in equilibrium with its vapor; $R_0=20$ $\mu$m). $10^4$ NPs with 500-nm-radius were always used, while three sets of the model parameters were applied. The red line -- the NP-NP interaction modeled with  Lennard-Jones potential only. The blue line -- apart from LJ interaction, half of the NPs were charged positively and another half -- negatively. The green line --  all the NPs had the same charge.}
    \label{fig:Df_evolution}
\end{figure}

Thus, it is generally possible to infer the evolution of the fractal dimension of an NP aggregate by analyzing the evolution of the observed light scattering patterns, at least for a small free aggregate. The evolution of $D_{\rm f}$, in turn, provides some information on the structure of the aggregate under scrutiny. Studying the possible evolution of the fractal dimension is therefore of great interest. In Fig. \ref{fig:LJ_only_SEM&Simulation} we show examples of NP aggregates of a mildly fractal structure: In the lower panel, there is an SEM image of an aggregate we obtained experimentally by slowly drying a microdroplet of SiO$_2$ NPs aqueous suspension with an addition of a surfactant (sodium dodecyl sulfate). In the upper panel, we present the simulation result of an aggregate formation in a non-evaporating droplet. The role of the surfactant was accounted for by modifying the interaction of NPs with the surface of the microdroplet only (a simplifying assumption). The temporal evolution of $D_{\rm f}$ for such parameters is presented in Fig. \ref{fig:Df_evolution} as the red curve. The blue and green curves in the figure also correspond to fairly realistic cases. In all three cases, the droplet was neither rotating nor evaporating, mimicking the bulk liquid's self-assembly conditions. Such conditions occur in a microdroplet in equilibrium with its vapor in a stable (stationary) environment. In the presented simulation, $10^4$ NPs with a radius of 500 nm were used. The specific parameters were: (i) The red line -- NP-NP interaction modeled with Lennard-Jones potential only -- both the dispersion medium and dispersed phase were non-polar. (ii) The blue line -- apart from LJ interaction, half of the NPs were charged positively and another half negatively -- two types of NPs of the same size, both the dispersion medium and NPs were polar. (iii) The green line -- LJ interaction, while all the NPs had the same charge -- a single type of NPs, both the dispersion medium and NPs were polar.

It should be pointed out that when an aggregate immersed in an evaporating droplet is considered, the 2D (surface) fractal dimension (evolution) also becomes of interest. When the properties of light scattering on such a composite microdroplet are sought, the droplet must be perceived as an optical (spherical) cavity, which provides a non-trivial illumination to the immersed aggregate. Due to evaporation, a train of morphology-dependent resonances (MDRs) is encountered. When the resonance condition is satisfied, most of the energy of the electromagnetic field is concentrated close to the microdroplet surface \cite{born2013principles}, especially when the condition for the so-called whispering gallery mode (WGM) is reached. It implies that the scattering of light by NPs in a microdroplet is often dominated by those lying closer to the surface. In consequence, it is mostly the surface structure of the (forming) aggregate that is reflected in scattering patterns. The concept of the surface-fractal is very similar to the mass-fractal and equations \ref{N} - \ref{r0} hold at lower dimensionality. However, the specifics of 2D geometry on a sphere must be accounted for.
\subsection{NPs distribution in the microdroplet}
Information provided by $D_f$ (evolution) alone can sometimes be ambiguous. Thus, the other parameter that we considered -- RDF -- describes how the NPs concentration varies as the function of distance from the chosen particle \cite{kopera2018computingRDFof_Particles}. It can be perceived as the probability of finding a particle pair separated by a distance $r$, normalized to a probability corresponding to a random distribution:
\begin{equation}
    g(r) = \sum_{i=1}^{N}\frac{\psi(r)_i/N}{(N-1)\delta V_r/V} \mbox{ ,}
\end{equation}
where $\psi(r)_i/N$ is the relative number of particles at a distance $r$ from an $i$th `trial particle' and $\delta V_r /V$ is the ratio of an `infinitesimal' volume at a distance $r$ (i.e. a spherical shell of thickness $\delta r$) to the total sample volume. The suspension is considered isotropic.

It is worth mentioning that RDF provides even more rigorous information on the light scattered by an aggregate (RDF being the Fourier transform of the so-called structure factor): 
\begin{equation}
    I_{\rm NPscat}(\vec{q}) = N_{\rm surf}^2\int e^{i\vec{q}\cdot\vec{r}}g(r)d\vec{r} \mbox{ .}
\end{equation}     
Generally, RDF describes the properties of suspension in bulk. In the case of a droplet of suspension, the influence of the sample's boundary should be considered and adequate normalization should be introduced. In the classical approach, the RDF function is averaged over the whole particle set (available volume) to get better statistics. In the case of an aggregate forming in the droplet, it is convenient to construct a normalized average NPs concentration as a function of the distance from the centre of the droplet $r_{\odot}$. We shall designate it by $\overline{C}_{{\rm NPs}}$: 
\begin{equation}
    \overline{C}_{{\rm NPs}}(r_{\odot}) = \frac{\psi(r_{\odot})/N}{\delta V_{r_{\odot}}/V} \mbox{ .}
\end{equation}
 Each of the introduced functions provides a somewhat different insight into the formation of the aggregate.
\section{Further examples of simulation and post-processing results}
Below, we present several successful conceptualizations of suspension microdroplet evolution obtained by post-processing the results of simulations. The final stage of modelled evolution was compared against the SEM images of dry NPs aggregates that had been obtained from experiments as described in section \ref{model_vs_SEM}. All the presented results but the last one were obtained using $10^4$ NPs. The last one used $4.1\times10^4$ NPs of two types. Apart from these, we performed several simulations with $2\times10^4$ NPs. The results of those -- movies, .bin, and parameters files -- can be found in the data repository \cite{20000inMendeleyData}.
\subsubsection{Aggregation in bulk, non-polar liquid}
\begin{figure}[htb]
    \centering
    \includegraphics[scale=0.3]{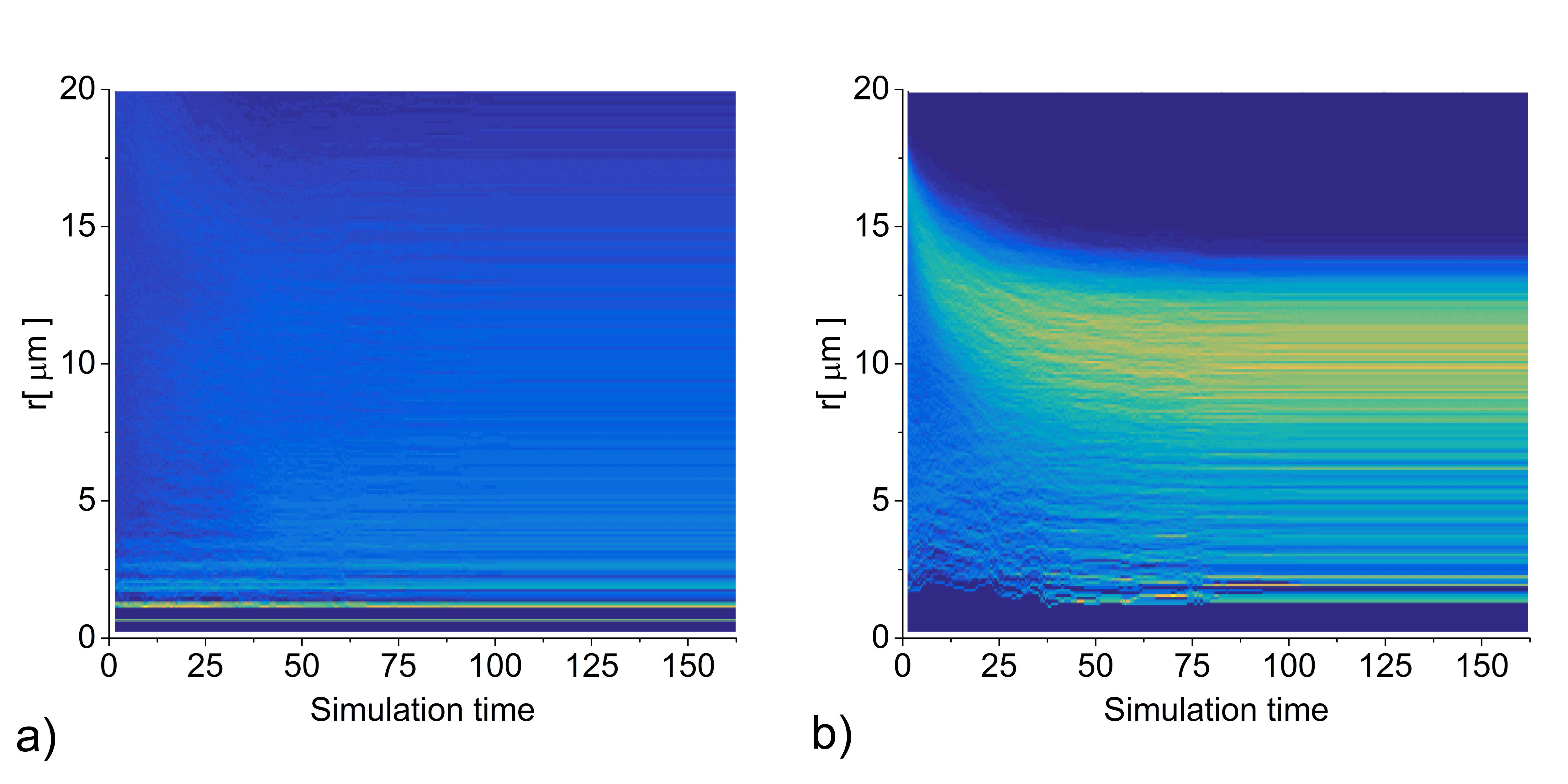}
    \caption{RDF functions' evolution obtained in post-processing for parameters corresponding to the red line in Fig. \ref{fig:Df_evolution} -- self-assembly in bulk liquid, only LJ interactions between NPs. Panel a) -- standard RDF, panel b) -- $\overline{C}_{{\rm NPs}}$}.
    \label{fig:RDF_Lennard_Jones}
\end{figure}

The first example of the application of both types of the distribution functions that we present in Fig. \ref{fig:RDF_Lennard_Jones} pertains to the temporal evolution of an NPs aggregate forming in the bulk non-polar liquid (mimicked by non-evaporating, stationary droplet). It corresponds to the red line in Fig. \ref{fig:Df_evolution} -- only LJ interactions between NPs are present; consistently $10^4$ NPs with 500-nm-radius were used. Panel a) shows the standard RDF, while panel b) -- $\overline{C}_{{\rm NPs}}$. In both panels, it can be seen that the initially randomly distributed nanoparticles start to form small clusters -- brighter, rather regularly spaced lines become more pronounced. The smaller, the more probable cluster, the brighter the line. Since the clusters are not linear, the higher-lying lines are spaced by fractions of the NP diameter and more diffuse. In panel a), only the isotropic ordering can be identified. On the other hand, since NPs were initially randomly spread over a finite volume, some evolution of their radial distribution versus their (initial) center of mass should be expected. And indeed, it is visualized in panel b). A movement of NPs towards their center of mass can be observed. A brighter region without a distinct pattern forms around 25 s, which signifies that NPs concentrate at a distance of $\sim$10 $\mu$m from their center of mass but do not form a regular shell.
\subsubsection{Aggregation in an evaporating droplet of non-polar liquid} 
\begin{figure}[htb]
    \centering
    \includegraphics[scale=0.61]{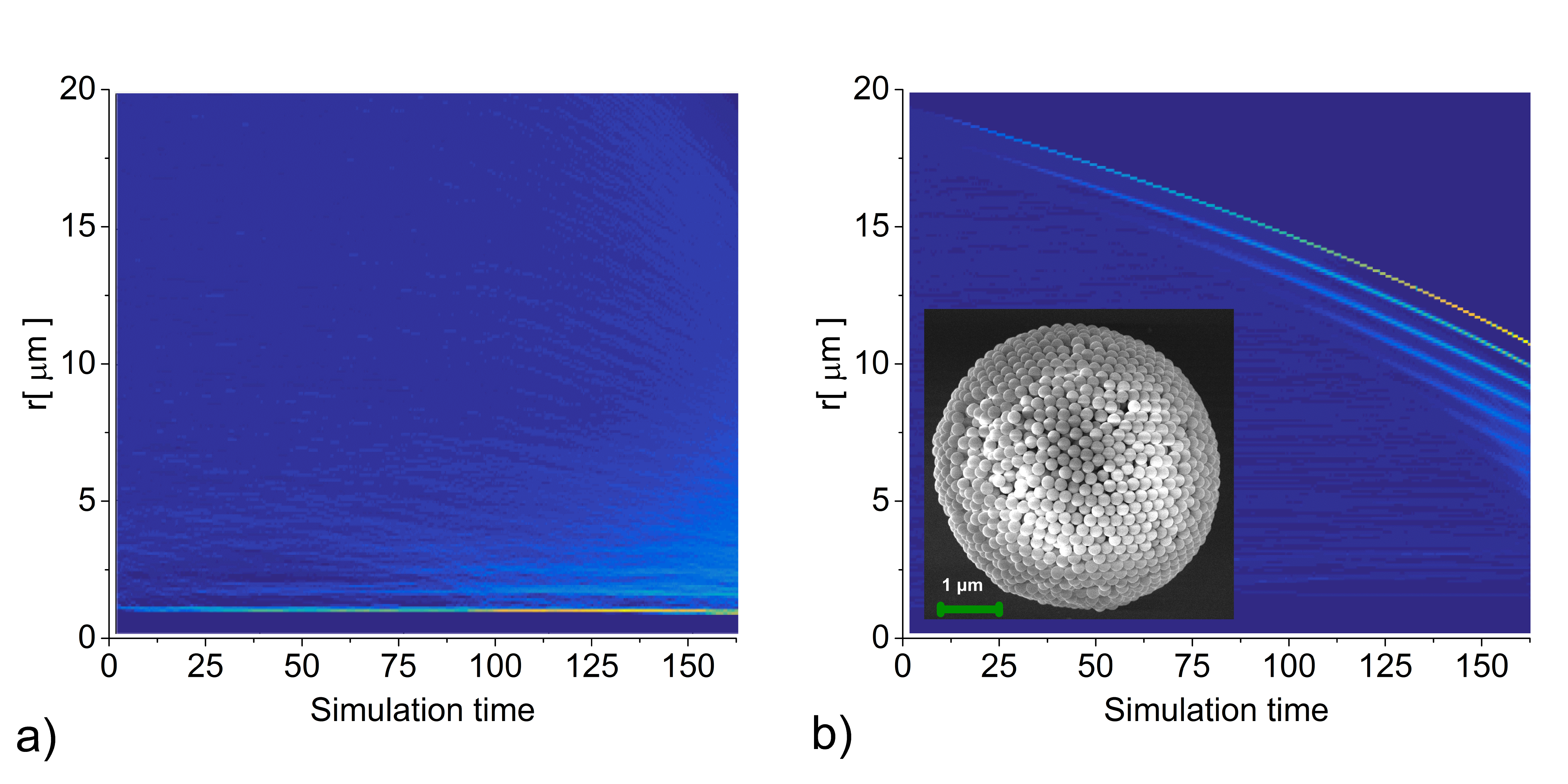}
    \caption{RDF (panel a)) and $\overline{C}_{{\rm NPs}}$ (panel b)) evolutions, obtained in post-processing, corresponding to the NPs aggregate formation in an evaporating microdroplet for the following model parameters: initial microdroplet radius $R_0=20$ $\mu$m, evaporation rate $A=3.5$, only LJ interactions between NPs ($\epsilon =0.023$, no NPs charge), no microdroplet rotation. Inset in panel b): SEM image of a dry aggregate of 125-nm-radius SiO$_2$ NPs, which we fabricated in the linear quadrupole trap. It corresponds to the end of the simulation -- right-hand side of both panels.}
    \label{fig:FRDF_01_Jun_2022}
\end{figure}

In contrast, Fig. \ref{fig:FRDF_01_Jun_2022} corresponds to aggregation in an evaporating droplet, while other parameters of the process were unchanged (only LJ interactions, no droplet rotation). From panel a) -- standard RDF -- we can only infer that the formation of long-range ordering is rather delayed. Only when we compare it with panel b) -- $\overline{C}_{{\rm NPs}}$ calculated relative to the center of the droplet -- the evolution of microdroplet/NPs aggregate morphology becomes more apparent. The process of building the successive NP layers at the droplet surface can be observed. The moving droplet boundary scavenges the NPs and pushes them towards the center until a dry NPs aggregate is formed. An SEM image of such NPs aggregate of 125-nm-radius SiO$_2$ NPs, which we fabricated in the linear quadrupole trap, is shown in the inset.
\subsubsection{Aggregation in a tumbling non-evaporating droplet of non-polar liquid}
\begin{figure}[htb]
    \centering
    \includegraphics[scale=0.3]{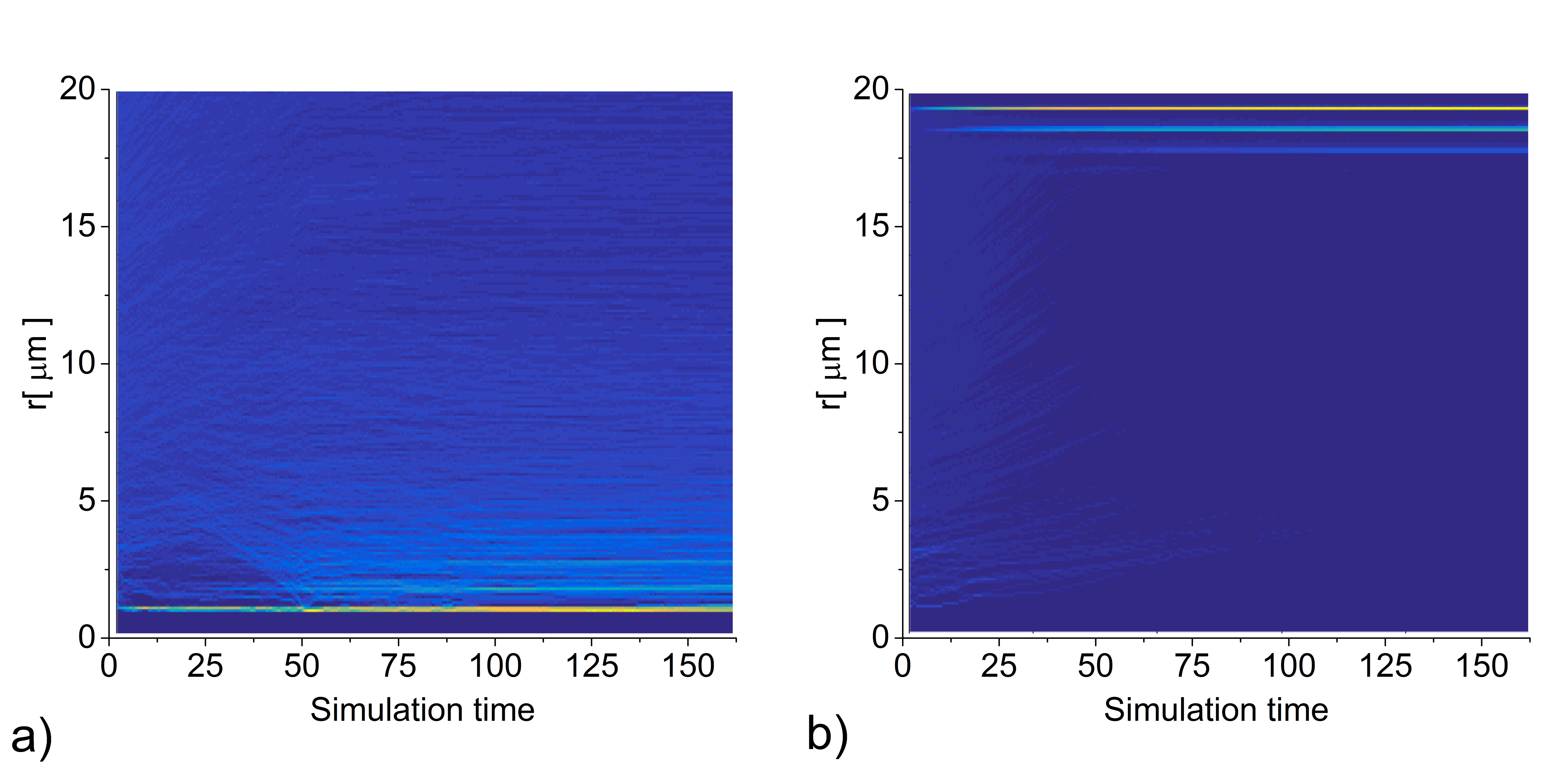}
    \caption{RDF (panel a)) and $\overline{C}_{{\rm NPs}}$ (panel b)) evolutions, obtained in post-processing, corresponding to the dry NPs aggregate formation presented in Fig. \ref{fig:Aggregate_evolution} (no microdroplet evaporation -- $R_0=20$ $\mu$m, $A=0$, random tumbling -- isotropic $\omega=1$ rad/s, only LJ interactions -- $\epsilon =0.023$).}
    \label{fig:FRDF_TwoType_q0_Rotation}
\end{figure}

A further example of the model's capabilities is presented in Fig.\ref{fig:FRDF_TwoType_q0_Rotation}, which corresponds to visualization in Fig. \ref{fig:Aggregate_evolution}. To remind: the aggregate was forming in a microdroplet that was not evaporating (was in equilibrium with the surrounding vapor) but tumbled randomly, which generated isotropic centrifugal force. Only LJ interactions were present, which yielded mainly a short-range ordering visible in panel a) as emerging, somewhat diffuse bright lines. The isotropic centrifugal force caused the formation of consecutive shells at the droplet surface, seen as bright lines at the top of panel b). The shells were densely packed (brightness/high contrast of the lines), which also manifests as a sharp bright line in panel a) at the distance of NP diameter from the test NP. 
\subsubsection{Aggregation in a rotating evaporating droplet of non-polar liquid}
\begin{figure}[htb]
    \centering
    \includegraphics[scale=0.5]{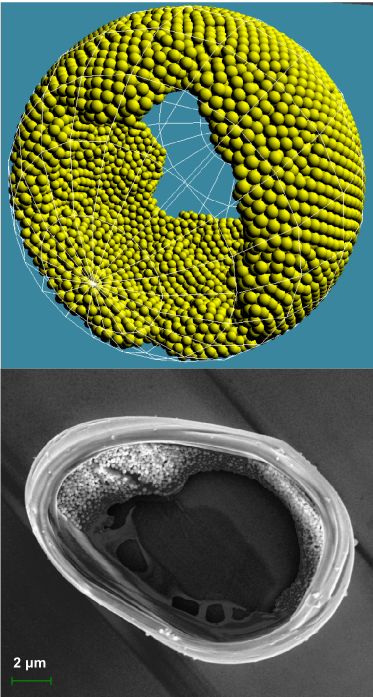}
    \caption{Examples of ring NPs aggregates. Lower panel -- SEM image of an NPs aggregate we obtained from drying a microdroplet of the aqueous suspension of 100-nm-radius SiO$_2$ NPs with an addition of sodium dodecyl sulfate (100 mM initially), in a linear electrodynamic quadrupole trap, for 45 min. The rotation conditions were not controlled. Upper panel -- the simulation result for the following model parameters: evaporation rate $A=1.5$, LJ interaction $\epsilon = 0.345$, no NP's charge, $\omega=2$ rad/s around the z-axis.}
    \label{fig:Oponka_SEM&Simulation}
\end{figure}

It must be however admitted that very similar distribution functions are obtained for a microdroplet/NPs aggregate rotating only around a single axis. The visualization of the simulation results together with the SEM image of a ring aggregate we produced in our physical experiment is shown in Fig. \ref{fig:Oponka_SEM&Simulation}. This illustrates the limitations of the distribution functions used and shows that, in general, the parametrization used for the aggregate description must be carefully selected.

\subsection{Aggregation of NPs of different sizes}
\begin{figure}
    \includegraphics[scale=0.4]{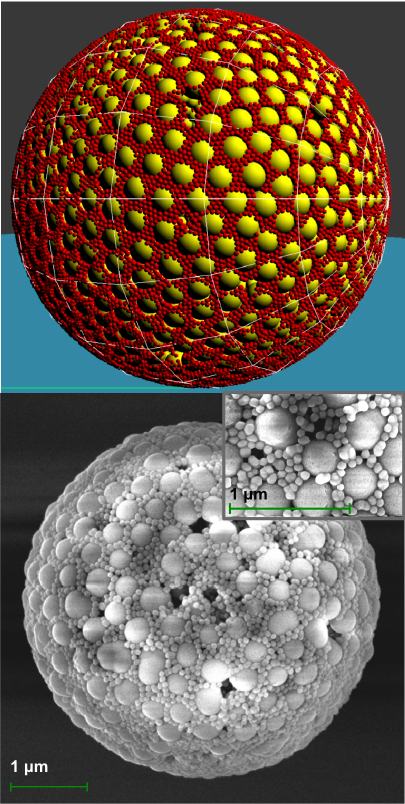}
    \includegraphics[scale=0.3]{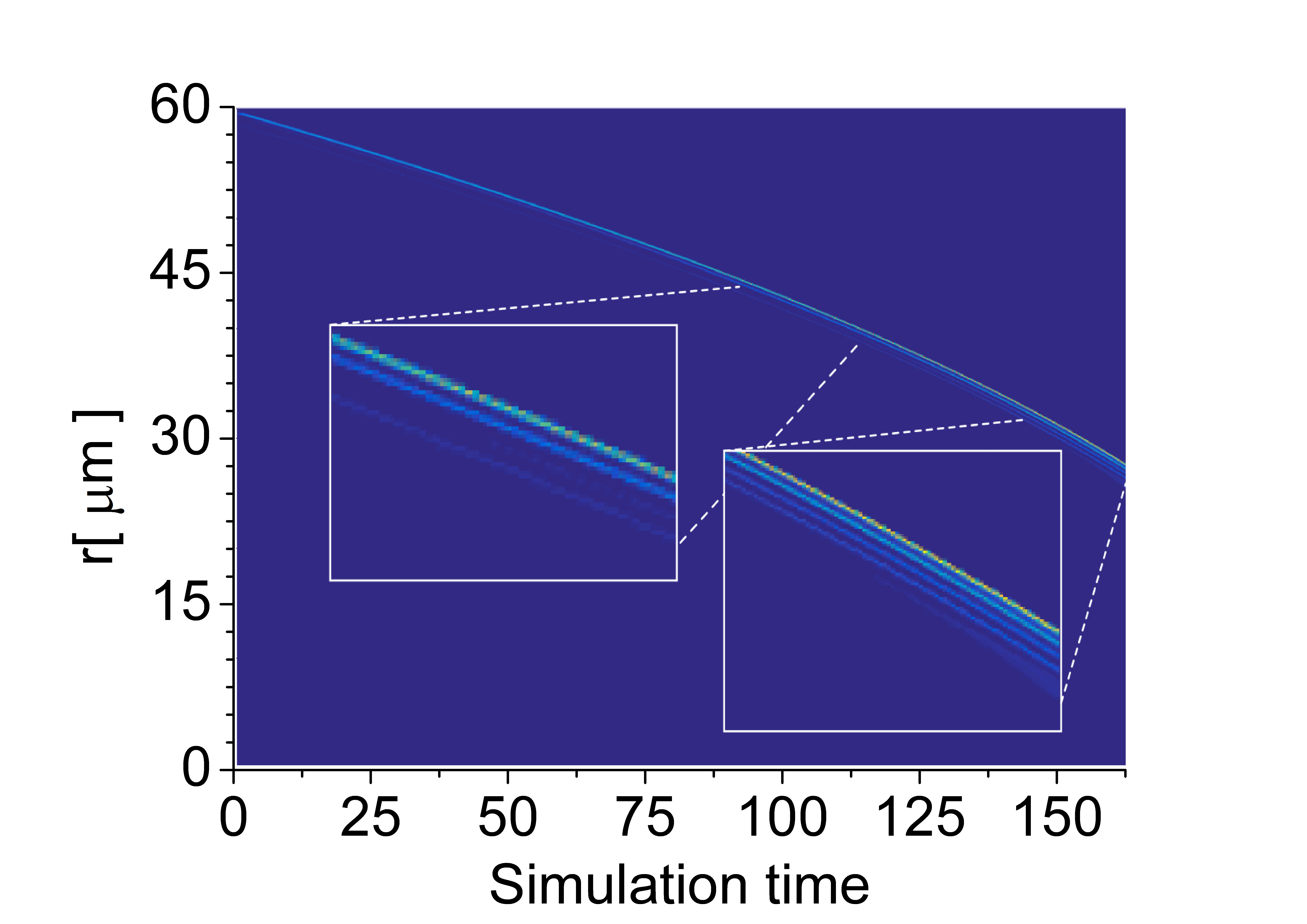}
    \caption{Examples of NPs aggregate composed of NPs with two significantly different diameters. Lower-left panel -- SEM image of an aggregate we obtained from drying a microdroplet of an aqueous suspension of 36 and 160-nm-radius SiO$_2$ NPs in 10:1 number ratio, in a linear electrodynamic quadrupole trap for $\sim2$ h. The inset corresponds to the higher SEM magnification. Upper-left panel -- the result of simulation for the following model parameters: initial microdroplet radius $R_0=60$ $\mu$m; evaporation rate $A=5$; random tumbling -- isotropic $\omega=1$ rad/s; NPs radius, number, and charge respectively: $r_{\rm big}=1.5$ nm, $N_{\rm big}=1500$, $q_{\rm big}=10$ and $r_{\rm small}=0.25$ nm, $N_{\rm small}=39500$, $q_{\rm small}=1$; LJ interactions between respective NPs: $\epsilon_{\rm big-big}=0.155$, $\epsilon_{\rm small-small}=0.155$; $\epsilon_{\rm big-small}=1.26$. The corresponding $\overline{C}_{{\rm NPs}}$ is shown in the right panel. The details of the function are shown in magnification in the insets.}
    \label{fig:Big&Small}
\end{figure}

As a final example, we present the result of a massive simulation of the evaporation of a microdroplet of a complex suspension containing NPs with two significantly different diameters. There were 39500 smaller and 1500 larger NPs. The simulation parameters were tuned to yield a relatively large hollow spherical aggregate at some stage of the evolution, as seen in the top-left panel of Fig. \ref{fig:Big&Small}. The aggregate in the SEM image in the lower-left panel is probably not quite hollow. We obtained it from a microdroplet of the aqueous suspension of 36 and 160-nm-radius SiO$_2$ NPs in a 10:1 number ratio. The numerical experiment enabled finding a possible set of parameters (in particular relative strengths of NP-NP interactions) that yielded mimicking of physical experiments findings. The $\overline{C}_{{\rm NPs}}$ presented in the right panel of Fig.\ref{fig:Big&Small} shows the formation of consecutive shells -- their structure is shown in magnification in the insets. The distribution of NPs of larger radii can be partly identified in the left inset as the less bright lines, which appear with some delay between the brighter lines.

\section{Conclusions}
The presented model has turned out very helpful for several applications. We see its major use for testing hypotheses and conceptualizing the evolution scenarios of evaporating microdroplets of suspensions. In this work, we've shown how it helped us to understand the phenomenon of scattering `revival', often observed during evaporation of composite microdroplets, and to visualize/mimic the structure of NPs aggregate that we produced by drying such microdroplets in our linear quadrupole trap. The conceptualization can be performed `at two speeds': First, with a larger time step, the evolution can be visualized in real time or even at increased speed. Key ideas and parameters can thus be quickly tested and failed hypotheses discarded. Next, the time step can be reduced to suppress possible harmful numerical effects, and the evolution is safely recorded for post-processing. We tested a few post-processing scripts and showed that such NPs aggregate properties like fractal dimension or radial distribution function -- routinely used in the context of light-scattering experiments -- could be easily obtained.

\section{Abbreviations}  
\begin{table}[H]
\begin{tabular}{l l}
 BD   & Brownian dynamics\\
 CFD  & computational fluid dynamics\\
 DLA  & diffusion-limited aggregation model\\
 DLCA & diffusion-limited cluster aggregation model\\
 GPU  & graphics processing unit\\ 
 GUI  & graphical user interface\\
 LAMMPS & large-scale atomic/molecular massively parallel simulator\\
 LBM  & lattice Boltzmann methods\\
 LD   & Langevin/Stokesian dynamics\\
 LJ   & Lenard-Jones (potential, force)\\
 MD   & molecular dynamics\\
 MDR  & morphology-dependent resonance\\
 NP   & nanoparticle\\ 
 RDF  & radial distribution functions\\
 RLA  & reaction-limited aggregation model\\
 RLCA & reaction-limited cluster aggregation model\\
 SEM  & scanning electron microscope\\ 
 WGM  & whispering gallery mode\\
 \end{tabular}
\end{table}

\section{Acknowledgment}
This research was funded in whole or in part by the National Science Centre, Poland, grant 2021/41/B/ST3/00069. For the purpose of Open Access, the authors has applied a CC-BY public copyright license to any Author Accepted Manuscript (AAM) version arising from this submission.






\begin{thebibliography}{55}
\expandafter\ifx\csname natexlab\endcsname\relax\def\natexlab#1{#1}\fi
\providecommand{\url}[1]{\texttt{#1}}
\providecommand{\href}[2]{#2}
\providecommand{\path}[1]{#1}
\providecommand{\DOIprefix}{doi:}
\providecommand{\ArXivprefix}{arXiv:}
\providecommand{\URLprefix}{URL: }
\providecommand{\Pubmedprefix}{pmid:}
\providecommand{\doi}[1]{\href{http://dx.doi.org/#1}{\path{#1}}}
\providecommand{\Pubmed}[1]{\href{pmid:#1}{\path{#1}}}
\providecommand{\bibinfo}[2]{#2}
\ifx\xfnm\relax \def\xfnm[#1]{\unskip,\space#1}\fi
\bibitem[{Tao and Moncrieff(2009)}]{tao2009multiscale}
\bibinfo{author}{W.-K. Tao}, \bibinfo{author}{M.~W. Moncrieff},
\newblock \bibinfo{title}{Multiscale cloud system modeling},
\newblock \bibinfo{journal}{Reviews of Geophysics} \bibinfo{volume}{47}
  (\bibinfo{year}{2009}). \DOIprefix\doi{10.1029/2008RG000276}.
\bibitem[{Byun(1999)}]{byun1999science}
\bibinfo{author}{D.~Byun},
\newblock \bibinfo{title}{Science algorithms of the epa models-3 community
  multi-scale air quality (cmaq) modeling system},
\newblock \bibinfo{journal}{EPA Report}  (\bibinfo{year}{1999}).
\bibitem[{Pruppacher et~al.(1998)Pruppacher, Klett, and
  Wang}]{pruppacher1998microphysics}
\bibinfo{author}{H.~R. Pruppacher}, \bibinfo{author}{J.~D. Klett},
  \bibinfo{author}{P.~K. Wang}, \bibinfo{title}{Microphysics of clouds and
  precipitation}, \bibinfo{publisher}{Taylor \& Francis}, \bibinfo{year}{1998}.
\bibitem[{Majeed and Wexler(2001)}]{majeed2001microphysics}
\bibinfo{author}{M.~A. Majeed}, \bibinfo{author}{A.~S. Wexler},
\newblock \bibinfo{title}{Microphysics of aqueous droplets in clouds and fogs
  as applied to pm-fine modeling},
\newblock \bibinfo{journal}{Atmospheric Environment} \bibinfo{volume}{35}
  (\bibinfo{year}{2001}) \bibinfo{pages}{1639--1653}.
  \DOIprefix\doi{10.1016/S1352-2310(00)00442-8}.
\bibitem[{Pani et~al.(2023)Pani, Lin, Wang, Chantara, Griffith, and
  Chang}]{pani2023aerosol}
\bibinfo{author}{S.~K. Pani}, \bibinfo{author}{N.-H. Lin},
  \bibinfo{author}{S.-H. Wang}, \bibinfo{author}{S.~Chantara},
  \bibinfo{author}{S.~M. Griffith}, \bibinfo{author}{J.~H.-W. Chang},
\newblock \bibinfo{title}{Aerosol mass scattering efficiencies and single
  scattering albedo under high mass loading in chiang mai valley, thailand},
\newblock \bibinfo{journal}{Atmospheric Environment} \bibinfo{volume}{308}
  (\bibinfo{year}{2023}) \bibinfo{pages}{119867}.
  \DOIprefix\doi{10.1016/j.atmosenv.2023.119867}.
\bibitem[{Ogren et~al.(1992)Ogren, Noone, Hallberg, Heintzenberg, Schell,
  Berner, Solly, Kruisz, Reischl, Arends et~al.}]{ogren1992measurements}
\bibinfo{author}{J.~Ogren}, \bibinfo{author}{K.~Noone},
  \bibinfo{author}{A.~Hallberg}, \bibinfo{author}{J.~Heintzenberg},
  \bibinfo{author}{D.~Schell}, \bibinfo{author}{A.~Berner},
  \bibinfo{author}{I.~Solly}, \bibinfo{author}{C.~Kruisz},
  \bibinfo{author}{G.~Reischl}, \bibinfo{author}{B.~Arends}, et~al.,
\newblock \bibinfo{title}{Measurements of the size dependence of the
  concentration of nonvolatile material in fog droplets},
\newblock \bibinfo{journal}{Tellus B} \bibinfo{volume}{44}
  (\bibinfo{year}{1992}) \bibinfo{pages}{570--580}.
  \DOIprefix\doi{10.1034/j.1600-0889.1992.t01-1-00010.x}.
\bibitem[{Gonzales and Murr(1977)}]{gonzales1977electron}
\bibinfo{author}{T.~Gonzales}, \bibinfo{author}{L.~Murr},
\newblock \bibinfo{title}{An electron microscopy study of particulates present
  in individual raindrops},
\newblock \bibinfo{journal}{Journal of Geophysical Research}
  \bibinfo{volume}{82} (\bibinfo{year}{1977}) \bibinfo{pages}{3161--3166}.
  \DOIprefix\doi{10.1029/JC082i021p03161}.
\bibitem[{Li et~al.(2013)Li, Wang, Collett~Jr, Chen, Zhang, Wang, and
  Wang}]{li2013microscopic}
\bibinfo{author}{W.~Li}, \bibinfo{author}{Y.~Wang}, \bibinfo{author}{J.~L.
  Collett~Jr}, \bibinfo{author}{J.~Chen}, \bibinfo{author}{X.~Zhang},
  \bibinfo{author}{Z.~Wang}, \bibinfo{author}{W.~Wang},
\newblock \bibinfo{title}{Microscopic evaluation of trace metals in cloud
  droplets in an acid precipitation region},
\newblock \bibinfo{journal}{Environmental science \& technology}
  \bibinfo{volume}{47} (\bibinfo{year}{2013}) \bibinfo{pages}{4172--4180}.
  \DOIprefix\doi{10.1021/es304779t}.
\bibitem[{Singh et~al.(2022)Singh, Srivastava, Pathak, and
  Shukla}]{singh2022quantifying}
\bibinfo{author}{A.~Singh}, \bibinfo{author}{A.~Srivastava},
  \bibinfo{author}{V.~Pathak}, \bibinfo{author}{A.~Shukla},
\newblock \bibinfo{title}{Quantifying the impact of biomass burning and dust
  storm activities on aerosol characteristics over the indo-gangetic basin},
\newblock \bibinfo{journal}{Atmospheric Environment} \bibinfo{volume}{270}
  (\bibinfo{year}{2022}) \bibinfo{pages}{118893}.
\bibitem[{Kocifaj and Bar{\'a}(2022)}]{kocifaj2022diffuse}
\bibinfo{author}{M.~Kocifaj}, \bibinfo{author}{S.~Bar{\'a}},
\newblock \bibinfo{title}{Diffuse light around cities: New perspectives in
  satellite remote sensing of nighttime aerosols},
\newblock \bibinfo{journal}{Atmospheric Research} \bibinfo{volume}{266}
  (\bibinfo{year}{2022}) \bibinfo{pages}{105969}.
\bibitem[{Brown et~al.(2012)Brown, Bonadonna, and Durant}]{brown2012review}
\bibinfo{author}{R.~Brown}, \bibinfo{author}{C.~Bonadonna},
  \bibinfo{author}{A.~Durant},
\newblock \bibinfo{title}{A review of volcanic ash aggregation},
\newblock \bibinfo{journal}{Physics and Chemistry of the Earth, Parts a/b/c}
  \bibinfo{volume}{45} (\bibinfo{year}{2012}) \bibinfo{pages}{65--78}.
\bibitem[{Xie et~al.(2014)Xie, Li, Li, Wang, Li, Chen, Li, and
  Xu}]{xie2014study}
\bibinfo{author}{Y.~Xie}, \bibinfo{author}{Z.~Li}, \bibinfo{author}{L.~Li},
  \bibinfo{author}{L.~Wang}, \bibinfo{author}{D.~Li},
  \bibinfo{author}{C.~Chen}, \bibinfo{author}{K.~Li}, \bibinfo{author}{H.~Xu},
\newblock \bibinfo{title}{Study on influence of different mixing rules on the
  aerosol components retrieval from ground-based remote sensing measurements},
\newblock \bibinfo{journal}{Atmospheric research} \bibinfo{volume}{145}
  (\bibinfo{year}{2014}) \bibinfo{pages}{267--278}.
\bibitem[{Jiaqiang et~al.(2022)Jiaqiang, Xu, Ma, Tan, Peng, Tan, and
  Chen}]{jiaqiang2022soot}
\bibinfo{author}{E.~Jiaqiang}, \bibinfo{author}{W.~Xu},
  \bibinfo{author}{Y.~Ma}, \bibinfo{author}{D.~Tan}, \bibinfo{author}{Q.~Peng},
  \bibinfo{author}{Y.~Tan}, \bibinfo{author}{L.~Chen},
\newblock \bibinfo{title}{Soot formation mechanism of modern automobile engines
  and methods of reducing soot emissions: A review},
\newblock \bibinfo{journal}{Fuel Processing Technology} \bibinfo{volume}{235}
  (\bibinfo{year}{2022}) \bibinfo{pages}{107373}.
\bibitem[{Santos et~al.(2015)Santos, Gabriel, Blanchy, Menes, Garc{\'i}a,
  Blanco, Arconada, and Neto}]{santosIndustrialApplicationsNanoparticles2015}
\bibinfo{author}{C.~S. Santos}, \bibinfo{author}{B.~Gabriel},
  \bibinfo{author}{M.~Blanchy}, \bibinfo{author}{O.~Menes},
  \bibinfo{author}{D.~Garc{\'i}a}, \bibinfo{author}{M.~Blanco},
  \bibinfo{author}{N.~Arconada}, \bibinfo{author}{V.~Neto},
\newblock \bibinfo{title}{Industrial {{Applications}} of {{Nanoparticles}}
  {\textendash} {{A Prospective Overview}}},
\newblock \bibinfo{journal}{Materials Today: Proceedings} \bibinfo{volume}{2}
  (\bibinfo{year}{2015}) \bibinfo{pages}{456--465}.
  \DOIprefix\doi{10.1016/j.matpr.2015.04.056}.
\bibitem[{Kumar et~al.(2021)Kumar, Verma, and Mandal}]{KUMAR2021109042}
\bibinfo{author}{N.~Kumar}, \bibinfo{author}{A.~Verma},
  \bibinfo{author}{A.~Mandal},
\newblock \bibinfo{title}{Formation, characteristics and oil industry
  applications of nanoemulsions: A review},
\newblock \bibinfo{journal}{Journal of Petroleum Science and Engineering}
  \bibinfo{volume}{206} (\bibinfo{year}{2021}) \bibinfo{pages}{109042}.
  \URLprefix
  \url{https://www.sciencedirect.com/science/article/pii/S0920410521006999}.
  \DOIprefix\doi{10.1016/j.petrol.2021.109042}.
\bibitem[{{Jafari Daghlian Sofla} et~al.(2018){Jafari Daghlian Sofla}, James,
  and Zhang}]{JAFARIDAGHLIANSOFLA2018559}
\bibinfo{author}{S.~{Jafari Daghlian Sofla}}, \bibinfo{author}{L.~A. James},
  \bibinfo{author}{Y.~Zhang},
\newblock \bibinfo{title}{Insight into the stability of hydrophilic silica
  nanoparticles in seawater for enhanced oil recovery implications},
\newblock \bibinfo{journal}{Fuel} \bibinfo{volume}{216} (\bibinfo{year}{2018})
  \bibinfo{pages}{559--571}. \URLprefix
  \url{https://www.sciencedirect.com/science/article/pii/S0016236117314941}.
  \DOIprefix\doi{10.1016/j.fuel.2017.11.091}.
\bibitem[{Sorensen(2001)}]{sorensen2001light}
\bibinfo{author}{C.~Sorensen},
\newblock \bibinfo{title}{Light scattering by fractal aggregates: a review},
\newblock \bibinfo{journal}{Aerosol Science \& Technology} \bibinfo{volume}{35}
  (\bibinfo{year}{2001}) \bibinfo{pages}{648--687}.
\bibitem[{Xu(2015)}]{xu2015light}
\bibinfo{author}{R.~Xu},
\newblock \bibinfo{title}{Light scattering: A review of particle
  characterization applications},
\newblock \bibinfo{journal}{Particuology} \bibinfo{volume}{18}
  (\bibinfo{year}{2015}) \bibinfo{pages}{11--21}.
\bibitem[{Ho{\l}yst et~al.(2013)Ho{\l}yst, Litniewski, Jakubczyk, Kolwas,
  Kolwas, Kowalski, Migacz, Palesa, and Zientara}]{holyst2013evaporation}
\bibinfo{author}{R.~Ho{\l}yst}, \bibinfo{author}{M.~Litniewski},
  \bibinfo{author}{D.~Jakubczyk}, \bibinfo{author}{K.~Kolwas},
  \bibinfo{author}{M.~Kolwas}, \bibinfo{author}{K.~Kowalski},
  \bibinfo{author}{S.~Migacz}, \bibinfo{author}{S.~Palesa},
  \bibinfo{author}{M.~Zientara},
\newblock \bibinfo{title}{Evaporation of freely suspended single droplets:
  experimental, theoretical and computational simulations},
\newblock \bibinfo{journal}{Reports on progress in physics}
  \bibinfo{volume}{76} (\bibinfo{year}{2013}) \bibinfo{pages}{034601}.
\bibitem[{Jakubczyk et~al.(2013)Jakubczyk, Derkachov, Kolwas, and
  Kolwas}]{jakubczykCombiningWeightingScatterometry2013}
\bibinfo{author}{D.~Jakubczyk}, \bibinfo{author}{G.~Derkachov},
  \bibinfo{author}{M.~Kolwas}, \bibinfo{author}{K.~Kolwas},
\newblock \bibinfo{title}{Combining weighting and scatterometry:
  {{Application}} to a levitated droplet of suspension},
\newblock \bibinfo{journal}{Journal of Quantitative Spectroscopy and Radiative
  Transfer} \bibinfo{volume}{126} (\bibinfo{year}{2013})
  \bibinfo{pages}{99--104}. \DOIprefix\doi{10.1016/j.jqsrt.2012.11.010}.
\bibitem[{Wozniak et~al.(2015)Wozniak, Derkachov, Kolwas, Archer,
  Wojciechowski, Jakubczyk, and Kolwas}]{wozniak2015formation}
\bibinfo{author}{M.~Wozniak}, \bibinfo{author}{G.~Derkachov},
  \bibinfo{author}{K.~Kolwas}, \bibinfo{author}{J.~Archer},
  \bibinfo{author}{T.~Wojciechowski}, \bibinfo{author}{D.~Jakubczyk},
  \bibinfo{author}{M.~Kolwas},
\newblock \bibinfo{title}{Formation of highly ordered spherical aggregates from
  drying microdroplets of colloidal suspension},
\newblock \bibinfo{journal}{Langmuir} \bibinfo{volume}{31}
  (\bibinfo{year}{2015}) \bibinfo{pages}{7860--7868}.
\bibitem[{Archer et~al.(2019)Archer, Kolwas, Wo{\'z}niak, Jakubczyk, Kolwas,
  Derkachov, and Wojciechowski}]{archer2019sodium}
\bibinfo{author}{J.~Archer}, \bibinfo{author}{M.~Kolwas},
  \bibinfo{author}{M.~Wo{\'z}niak}, \bibinfo{author}{D.~Jakubczyk},
  \bibinfo{author}{K.~Kolwas}, \bibinfo{author}{G.~Derkachov},
  \bibinfo{author}{T.~Wojciechowski},
\newblock \bibinfo{title}{Sodium dodecyl sulfate microaggregates with diversely
  developed surfaces: Formation from free microdroplets of colloidal
  suspension},
\newblock \bibinfo{journal}{The European Physical Journal Plus}
  \bibinfo{volume}{134} (\bibinfo{year}{2019}) \bibinfo{pages}{1--12}.
\bibitem[{Dey et~al.(2024)Dey, Mallen, Stone, and Joshi}]{dey2024using}
\bibinfo{author}{S.~Dey}, \bibinfo{author}{E.~Mallen},
  \bibinfo{author}{B.~Stone}, \bibinfo{author}{Y.~Joshi},
\newblock \bibinfo{title}{Using multiscale atmospheric modeling to explore the
  impact of surface albedo on anthropogenic heat release},
\newblock \bibinfo{journal}{ASME Journal of Heat and Mass Transfer}
  \bibinfo{volume}{146} (\bibinfo{year}{2024}).
  \DOIprefix\doi{10.1115/1.4065088}.
\bibitem[{Chen et~al.(2023)Chen, Zhao, Shen, and Fan}]{chen2023influence}
\bibinfo{author}{A.~Chen}, \bibinfo{author}{C.~Zhao},
  \bibinfo{author}{L.~Shen}, \bibinfo{author}{T.~Fan},
\newblock \bibinfo{title}{Influence of aerosol properties and surface albedo on
  radiative forcing efficiency of key aerosol types using global aeronet data},
\newblock \bibinfo{journal}{Atmospheric Research} \bibinfo{volume}{282}
  (\bibinfo{year}{2023}) \bibinfo{pages}{106519}.
  \DOIprefix\doi{10.1016/j.atmosres.2022.106519}.
\bibitem[{Herman and Browning(1975)}]{herman1975effect}
\bibinfo{author}{B.~M. Herman}, \bibinfo{author}{S.~R. Browning},
\newblock \bibinfo{title}{The effect of aerosols on the earth-atmosphere
  albedo},
\newblock \bibinfo{journal}{Journal of Atmospheric Sciences}
  \bibinfo{volume}{32} (\bibinfo{year}{1975}) \bibinfo{pages}{1430--1445}.
  \DOIprefix\doi{10.1175/1520-0469(1975)032<1430:TEOAOT>2.0.CO;2}.
\bibitem[{Takemura et~al.(2002)Takemura, Nakajima, Dubovik, Holben, and
  Kinne}]{takemura2002single}
\bibinfo{author}{T.~Takemura}, \bibinfo{author}{T.~Nakajima},
  \bibinfo{author}{O.~Dubovik}, \bibinfo{author}{B.~N. Holben},
  \bibinfo{author}{S.~Kinne},
\newblock \bibinfo{title}{Single-scattering albedo and radiative forcing of
  various aerosol species with a global three-dimensional model},
\newblock \bibinfo{journal}{Journal of Climate} \bibinfo{volume}{15}
  (\bibinfo{year}{2002}) \bibinfo{pages}{333--352}.
  \DOIprefix\doi{10.1175/1520-0442(2002)015<0333:SSAARF>2.0.CO;2}.
\bibitem[{Chen et~al.(2024)Chen, Zhao, Zhang, Yang, and Li}]{chen2024surface}
\bibinfo{author}{A.~Chen}, \bibinfo{author}{C.~Zhao},
  \bibinfo{author}{H.~Zhang}, \bibinfo{author}{Y.~Yang},
  \bibinfo{author}{J.~Li},
\newblock \bibinfo{title}{Surface albedo regulates aerosol direct climate
  effect},
\newblock \bibinfo{journal}{Nature Communications} \bibinfo{volume}{15}
  (\bibinfo{year}{2024}) \bibinfo{pages}{7816}.
  \DOIprefix\doi{10.1038/s41467-024-52255-z}.
\bibitem[{Weitkamp et~al.(2005)}]{weitkamp2005range}
\bibinfo{author}{C.~Weitkamp}, et~al.,
\newblock \bibinfo{title}{Range-resolved optical remote sensing of the
  atmosphere},
\newblock \bibinfo{journal}{Springer-Verlag New York} \bibinfo{volume}{102}
  (\bibinfo{year}{2005}) \bibinfo{pages}{241--303}.
\bibitem[{Kong et~al.(2022)Kong, Sato, and Bi}]{kong2022lidar}
\bibinfo{author}{S.~Kong}, \bibinfo{author}{K.~Sato}, \bibinfo{author}{L.~Bi},
\newblock \bibinfo{title}{Lidar ratio--depolarization ratio relations of
  atmospheric dust aerosols: The super-spheroid model and high spectral
  resolution lidar observations},
\newblock \bibinfo{journal}{Journal of Geophysical Research: Atmospheres}
  \bibinfo{volume}{127} (\bibinfo{year}{2022}) \bibinfo{pages}{e2021JD035629}.
\bibitem[{Li et~al.(2020)Li, Di, Wang, Han, Hua, and Li}]{li2020retrieval}
\bibinfo{author}{S.~Li}, \bibinfo{author}{H.~Di}, \bibinfo{author}{Q.~Wang},
  \bibinfo{author}{G.~Han}, \bibinfo{author}{D.~Hua}, \bibinfo{author}{Y.~Li},
\newblock \bibinfo{title}{Retrieval of the aerosol extinction coefficient of
  1064nm based on high-spectral-resolution lidar},
\newblock \bibinfo{journal}{Journal of Quantitative Spectroscopy and Radiative
  Transfer} \bibinfo{volume}{256} (\bibinfo{year}{2020})
  \bibinfo{pages}{107298}.
\bibitem[{Garbacz(1978)}]{garbacz1978chaff}
\bibinfo{author}{R.~Garbacz},
\newblock \bibinfo{title}{Chaff radar cross section studies and calculations},
\newblock \bibinfo{journal}{Interim Report, Ohio State Univ. Columbus
  Electroscience Lab}  (\bibinfo{year}{1978}).
\bibitem[{Witten~Jr and Sander(1981)}]{witten1981diffusion}
\bibinfo{author}{T.~A. Witten~Jr}, \bibinfo{author}{L.~M. Sander},
\newblock \bibinfo{title}{Diffusion-limited aggregation, a kinetic critical
  phenomenon},
\newblock \bibinfo{journal}{Physical review letters} \bibinfo{volume}{47}
  (\bibinfo{year}{1981}) \bibinfo{pages}{1400}.
\bibitem[{{L.M.
  Sander}(2000)}]{l.m.sanderDiffusionlimitedAggregationKinetic2000}
\bibinfo{author}{{L.M. Sander}},
\newblock \bibinfo{title}{Diffusion-limited aggregation: {{A}} kinetic critical
  phenomenon?},
\newblock \bibinfo{journal}{Contemporary Physics} \bibinfo{volume}{41}
  (\bibinfo{year}{2000}) \bibinfo{pages}{203--218}.
  \DOIprefix\doi{10.1080/001075100409698}.
\bibitem[{Kolb et~al.(1983)Kolb, Botet, and
  Jullien}]{kolbScalingKineticallyGrowing1983}
\bibinfo{author}{M.~Kolb}, \bibinfo{author}{R.~Botet},
  \bibinfo{author}{R.~Jullien},
\newblock \bibinfo{title}{Scaling of {{Kinetically Growing Clusters}}},
\newblock \bibinfo{journal}{Physical Review Letters} \bibinfo{volume}{51}
  (\bibinfo{year}{1983}) \bibinfo{pages}{1123--1126}.
  \DOIprefix\doi{10.1103/PhysRevLett.51.1123}.
\bibitem[{Meakin et~al.(1986)Meakin, Ramanlal, Sander, and
  Ball}]{meakinBallisticDepositionSurfaces1986}
\bibinfo{author}{P.~Meakin}, \bibinfo{author}{P.~Ramanlal},
  \bibinfo{author}{L.~M. Sander}, \bibinfo{author}{R.~C. Ball},
\newblock \bibinfo{title}{Ballistic deposition on surfaces},
\newblock \bibinfo{journal}{Physical Review A} \bibinfo{volume}{34}
  (\bibinfo{year}{1986}) \bibinfo{pages}{5091--5103}.
  \DOIprefix\doi{10.1103/PhysRevA.34.5091}.
\bibitem[{Meakin and Family(1988)}]{meakinStructureKineticsReactionlimited1988}
\bibinfo{author}{P.~Meakin}, \bibinfo{author}{F.~Family},
\newblock \bibinfo{title}{Structure and kinetics of reaction-limited
  aggregation},
\newblock \bibinfo{journal}{Physical Review A} \bibinfo{volume}{38}
  (\bibinfo{year}{1988}) \bibinfo{pages}{2110--2123}.
  \DOIprefix\doi{10.1103/PhysRevA.38.2110}.
\bibitem[{Ulberg et~al.(1993)Ulberg, Churaev, Ilyin, and
  Malashenko}]{ulbergMolecularDynamicsSimulation1993}
\bibinfo{author}{D.~Ulberg}, \bibinfo{author}{N.~Churaev},
  \bibinfo{author}{V.~Ilyin}, \bibinfo{author}{G.~Malashenko},
\newblock \bibinfo{title}{Molecular dynamics simulation of the aggregation of
  colloidal particles},
\newblock \bibinfo{journal}{Colloids and Surfaces A: Physicochemical and
  Engineering Aspects} \bibinfo{volume}{80} (\bibinfo{year}{1993})
  \bibinfo{pages}{93--102}. \DOIprefix\doi{10.1016/0927-7757(93)80186-I}.
\bibitem[{Markutsya et~al.(2008)Markutsya, Subramaniam, Vigil, and
  Fox}]{markutsyaBrownianDynamicsSimulation2008}
\bibinfo{author}{S.~Markutsya}, \bibinfo{author}{S.~Subramaniam},
  \bibinfo{author}{R.~D. Vigil}, \bibinfo{author}{R.~O. Fox},
\newblock \bibinfo{title}{On {{Brownian Dynamics Simulation}} of {{Nanoparticle
  Aggregation}}},
\newblock \bibinfo{journal}{Industrial \& Engineering Chemistry Research}
  \bibinfo{volume}{47} (\bibinfo{year}{2008}) \bibinfo{pages}{3338--3345}.
  \DOIprefix\doi{10.1021/ie0711168}.
\bibitem[{{Garc{\'i}a-Palacios} and
  L{\'a}zaro(1998)}]{garcia-palaciosLangevindynamicsStudyDynamical1998}
\bibinfo{author}{J.~L. {Garc{\'i}a-Palacios}}, \bibinfo{author}{F.~J.
  L{\'a}zaro},
\newblock \bibinfo{title}{Langevin-dynamics study of the dynamical properties
  of small magnetic particles},
\newblock \bibinfo{journal}{Physical Review B} \bibinfo{volume}{58}
  (\bibinfo{year}{1998}) \bibinfo{pages}{14937--14958}.
  \DOIprefix\doi{10.1103/PhysRevB.58.14937}.
\bibitem[{Wendt et~al.(2008)Wendt, Anderson, and {Von Karman Institute for
  Fluid Dynamics}}]{wendtComputationalFluidDynamics2008}
\bibinfo{editor}{J.~F. Wendt}, \bibinfo{editor}{J.~D. Anderson},
  \bibinfo{editor}{{Von Karman Institute for Fluid Dynamics}} (Eds.),
  \bibinfo{title}{Computational Fluid Dynamics: An Introduction},
  \bibinfo{edition}{3rd ed} ed., \bibinfo{publisher}{{Springer}},
  \bibinfo{address}{{Berlin ; [London]}}, \bibinfo{year}{2008}.
\bibitem[{Green(2010)}]{green2010particle}
\bibinfo{author}{S.~Green},
\newblock \bibinfo{title}{Particle simulation using cuda},
\newblock \bibinfo{journal}{NVIDIA whitepaper} \bibinfo{volume}{6}
  (\bibinfo{year}{2010}) \bibinfo{pages}{121--128}.
\bibitem[{Derkachov et~al.(2008)Derkachov, Kolwas, Jakubczyk, Zientara, and
  Kolwas}]{derkachov2008drying}
\bibinfo{author}{G.~Derkachov}, \bibinfo{author}{K.~Kolwas},
  \bibinfo{author}{D.~Jakubczyk}, \bibinfo{author}{M.~Zientara},
  \bibinfo{author}{M.~Kolwas},
\newblock \bibinfo{title}{Drying of a microdroplet of water suspension of
  nanoparticles: from surface aggregates to microcrystal},
\newblock \bibinfo{journal}{The Journal of Physical Chemistry C}
  \bibinfo{volume}{112} (\bibinfo{year}{2008}) \bibinfo{pages}{16919--16923}.
\bibitem[{G.(2008)}]{MatlabSimulinkAggregation}
\bibinfo{author}{G.~Derkachov}, \bibinfo{title}{Matlab/simulink model of aggregation},
  \bibinfo{year}{2008}. \URLprefix
  \url{https://github.com/Trankwery/Aggregation\_Model\_Matlab\_Simulink\_Code}.
\bibitem[{T.(2023)}]{githubGPUparticles}
\bibinfo{author}{T.~Jakubczyk}, \bibinfo{title}{Particles}, \bibinfo{year}{2023}.
  \URLprefix \url{https://github.com/Trankwery/particles}.
\bibitem[{NVIDIA(2017)}]{CUDA}
\bibinfo{author}{NVIDIA}, \bibinfo{title}{{NVIDIA} cuda toolkit 8.0},
\bibinfo{year}{2017}
  \URLprefix \url{https://developer.nvidia.com/cuda-80-ga2-download-archive}.
\bibitem[{Jakubczyk et~al.(2012)Jakubczyk, Kolwas, Derkachov, Kolwas, and
  Zientara}]{jakubczyk2012evaporation}
\bibinfo{author}{D.~Jakubczyk}, \bibinfo{author}{M.~Kolwas},
  \bibinfo{author}{G.~Derkachov}, \bibinfo{author}{K.~Kolwas},
  \bibinfo{author}{M.~Zientara},
\newblock \bibinfo{title}{Evaporation of micro-droplets: the"
  radius-square-law" revisited},
\newblock \bibinfo{journal}{Acta Physica Polonica A} \bibinfo{volume}{122}
  (\bibinfo{year}{2012}) \bibinfo{pages}{709--716}.
\bibitem[{Ferri et~al.(2022)Ferri, Humbert, Schweitzer, Digne, Lefebvre, and
  Moreaud}]{ferri2022MassFractal}
\bibinfo{author}{G.~Ferri}, \bibinfo{author}{S.~Humbert},
  \bibinfo{author}{J.~Schweitzer}, \bibinfo{author}{M.~Digne},
  \bibinfo{author}{V.~Lefebvre}, \bibinfo{author}{M.~Moreaud},
\newblock \bibinfo{title}{Mass fractal dimension from 2d microscopy images via
  an aggregation model with variable compactness},
\newblock \bibinfo{journal}{Journal of Microscopy} \bibinfo{volume}{286}
  (\bibinfo{year}{2022}) \bibinfo{pages}{31--41}.
\bibitem[{Larsen and Shaw(2018)}]{larsen2018methodComputingRDF_Cloud}
\bibinfo{author}{M.~Larsen}, \bibinfo{author}{R.~Shaw},
\newblock \bibinfo{title}{A method for computing the three-dimensional radial
  distribution function of cloud particles from holographic images},
\newblock \bibinfo{journal}{Atmospheric Measurement Techniques}
  \bibinfo{volume}{11} (\bibinfo{year}{2018}) \bibinfo{pages}{4261--4272}.
  \DOIprefix\doi{10.5194/amt-11-4261-2018}.
\bibitem[{Onofri et~al.(2011)Onofri, Wozniak, and
  Barbosa}]{onofriOpticalCharacterisationNanoparticle2011}
\bibinfo{author}{F.~Onofri}, \bibinfo{author}{M.~Wozniak},
  \bibinfo{author}{S.~Barbosa},
\newblock \bibinfo{title}{On the {{Optical Characterisation}} of
  {{Nanoparticle}} and their {{Aggregates}} in {{Plasma Systems}}},
\newblock \bibinfo{journal}{Contributions to Plasma Physics}
  \bibinfo{volume}{51} (\bibinfo{year}{2011}) \bibinfo{pages}{228--236}.
  \DOIprefix\doi{10.1002/ctpp.201000056}.
\bibitem[{Jungblut and
  Eychm{\"u}ller(2019)}]{jungblutModelingNanoparticleAggregation2019}
\bibinfo{author}{S.~Jungblut}, \bibinfo{author}{A.~Eychm{\"u}ller},
  \bibinfo{title}{Modeling Nanoparticle Aggregation},
  volume~\bibinfo{volume}{15}, \bibinfo{publisher}{{Royal Society of
  Chemistry}}, \bibinfo{address}{{Cambridge}}, \bibinfo{year}{2019}.
  \DOIprefix\doi{10.1039/9781788015868-00001}.
\bibitem[{Kopera and Retsch(2018)}]{kopera2018computingRDFof_Particles}
\bibinfo{author}{B.~Kopera}, \bibinfo{author}{M.~Retsch},
\newblock \bibinfo{title}{Computing the 3d radial distribution function from
  particle positions: an advanced analytic approach},
\newblock \bibinfo{journal}{Analytical chemistry} \bibinfo{volume}{90}
  (\bibinfo{year}{2018}) \bibinfo{pages}{13909--13914}.
  \DOIprefix\doi{10.1021/acs.analchem.8b03157}.
\bibitem[{Lapuerta et~al.(2010)Lapuerta, Martos, and
  {Mart{\'i}n-Gonz{\'a}lez}}]{lapuertaGeometricalDeterminationLacunarity2010}
\bibinfo{author}{M.~Lapuerta}, \bibinfo{author}{F.~Martos},
  \bibinfo{author}{G.~{Mart{\'i}n-Gonz{\'a}lez}},
\newblock \bibinfo{title}{Geometrical determination of the lacunarity of
  agglomerates with integer fractal dimension},
\newblock \bibinfo{journal}{Journal of Colloid and Interface Science}
  \bibinfo{volume}{346} (\bibinfo{year}{2010}) \bibinfo{pages}{23--31}.
  \DOIprefix\doi{10.1016/j.jcis.2010.02.016}.
\bibitem[{Mroczka et~al.(2012)Mroczka, Wo{\'z}niak, and
  Onofri}]{mroczka2012AlgorithmsMethodsOpticalStructureFactor}
\bibinfo{author}{J.~Mroczka}, \bibinfo{author}{M.~Wo{\'z}niak},
  \bibinfo{author}{F.~Onofri},
\newblock \bibinfo{title}{Algorithms and methods for analysis of the optical
  structure factor of fractal aggregates},
\newblock \bibinfo{journal}{Metrology and Measurement Systems}
  \bibinfo{volume}{19} (\bibinfo{year}{2012}) \bibinfo{pages}{459--470}.
\bibitem[{M. and E.(2013)}]{born2013principles}
\bibinfo{author}{B.~M.}, \bibinfo{author}{W.~E.}, \bibinfo{title}{Principles of
  optics: electromagnetic theory of propagation, interference and diffraction
  of light}, \bibinfo{publisher}{Elsevier}, \bibinfo{year}{2013}.
\bibitem[{Derkachov et~al.(2023)Derkachov, Alikhanzadeh-Arani, Jakubczyk, and
  Jakubczyk}]{20000inMendeleyData}
\bibinfo{author}{G.~Derkachov}, \bibinfo{author}{S.~Alikhanzadeh-Arani},
  \bibinfo{author}{T.~Jakubczyk}, \bibinfo{author}{D.~Jakubczyk},
  \bibinfo{title}{The results of md-like simulations of the self-assembly of 20
  000 nanoparticles}, \bibinfo{year}{2023}. \URLprefix
  \url{https://data.mendeley.com/datasets/6f5g4df4kh}.
  \DOIprefix\doi{10.17632/6f5g4df4kh.1}.

\end{thebibliography}


\end{document}